\documentclass[12pt]{iopart}

\usepackage{graphicx}
\usepackage{subfig}
\usepackage{float}
\usepackage{times}
\usepackage{natbib}
\newcommand{\enzo}{\small{ENZO}}

\begin{document}

\title[Simulations of extragalactic magnetic fields]{Simulations of extragalactic magnetic fields and of their observables.}

\author{F. Vazza $^{1,2,3}$, M. Br\"{u}ggen $^{2}$, C. Gheller$^{4}$, S. Hackstein$^{2}$, D. Wittor$^{2}$, P. M. Hinz$^{2}$}

\address{
$^{1}$ University of Bologna, Department of Physics and Astronomy, Via Gobetti 93/2, I-40129, Bologna, Italy;\\
  $^{2}$ Universit\"{a}t Hamburg, Hamburger Sternwarte, Gojenbergsweg 112, 40129, Hamburg, Germany;\\
 $^{3}$ Istituto di Radioastronomia, INAF, Via Gobetti 101, 40129 Bologna, Italy;\\
  $^{4}$ CSCS-ETHZ, Via Trevano 131, Lugano, Switzerland.
}

\ead{vazza@ira.inaf.it}
\vspace{10pt}
\begin{indented}
\item[]February 2017
\end{indented}

\begin{abstract}
  The origin of extragalactic magnetic fields is still poorly understood.  Based on a dedicated suite of cosmological magneto-hydrodynamical simulations with the {\enzo} code we have performed a survey of different models that may have caused present-day magnetic fields in galaxies and galaxy clusters. The outcomes of these models differ in cluster outskirts, filaments, sheets and voids and we use these simulations to find observational signatures of magnetogenesis. With these simulations, we predict the signal of extragalactic magnetic fields in radio observations of synchrotron emission from the cosmic web, in Faraday Rotation, in the propagation of Ultra High Energy Cosmic Rays, in the polarized signal from Fast Radio Bursts at cosmological distance  and in  spectra of distant blazars.
In general, {\it primordial} scenarios in which present-day magnetic fields originate from the amplification of weak ($\leq ~\rm nG$) uniform seed fields result  more homogeneous and relatively easier to observe magnetic fields than than {\it astrophysical} scenarios, in which
present-day fields are the product of feedback processes triggered by stars and active galaxies. In the near future the best evidence for the origin of cosmic magnetic fields will most likely come from a combination of synchrotron emission and Faraday Rotation observed at the periphery of large-scale structures. 

\end{abstract}

\section{Introduction}

Despite the progress in measuring magnetic fields across cosmic scales in the last few decades, the origin of extragalactic magnetic fields is still a mystery \citep[e.g.][]{2002RvMP...74..775W, 2012SSRv..166....1R,wi11}. We understand by and large how the inhomogeneities of the cosmic microwave background (CMB) at $z \sim 1100$ are related to the distribution of dark and baryonic matter, but we know little about how magnetic fields in galaxies and clusters of galaxies have formed. It is generally believed that today's magnetic fields in galaxies \citep[e.g.][for a few recent works on the subject]{beck12,2014ApJ...783L..20P,2016MNRAS.457.1722R,2017arXiv170405845R} or in the intracluster medium \citep[e.g.][]{do99,br05,ry08,xu09,bm16} must have been the result of amplification of so-called seed fields. But are the  seed  fields already in place at the epoch of the CMB and just too weak to 
be detected? Or did the seed fields arise during galaxy formation, and were powered by magnetized winds and jets from compact objects?
These are tough questions since observations of magnetic fields in rarefied environment are scarce and their interpretation is difficult. 
 
On one hand, the discovery of primordial magnetic fields will inspire interesting physics of the early Universe that may involve the generation of currents during inflation, phase transitions and baryogenesis \citep[e.g.][]{1973MNRAS.165..185H,2010PhRvD..82h3005K,wi11,2011ApJ...726...78K,2013A&ARv..21...62D,sub16,2016PhyS...91j4008K}.
Depending on the particular magnetization mechanism at play, these primordial seed fields may be characterized by small ($\leq ~\rm Mpc$, e.g. \citealt{1967SvA....10..634C}) or large \citep[e.g.][]{1970SvA....13..608Z,1988PhRvD..37.2743T} coherence lengths, whose structure may still persist in todays voids, possibly also carrying information on the generation of primordial helicity \citep[e.g.][]{2005A&A...433L..53S,2009IJMPD..18.1395C,2016PhyS...91j4008K}.
On the other hand, if magnetic fields have been released by processes triggered during galaxy formation, they might have affected the transport of heat, entropy, metals and cosmic rays in forming cosmic structures \citep[e.g.][]{2016mssf.book...93P}. 
Additional processes such as the ``Biermann-battery'' mechanism \citep[][]{1997ApJ...480..481K}, aperiodic plasma fluctuations in the inter-galactic plasma \citep[][]{2012ApJ...758..102S},  resistive mechanisms \citep[][]{mb11} or ionization fronts around the first stars  \citep[][]{2005A&A...443..367L} might provide additional amplification to the primordial fields.  Overall, the present uncertainty in primordial magnetic fields is uncomfortably large, allowing for fields in the range of $\sim ~10^{-34}-10^{-10}$ G.\\

Inside cosmic structures, weak seed fields can be amplified by a small-scale dynamo triggered by the turbulence induced by mergers and accretions \citep[][]{ry08,2011ApJ...731...62F,2013NJPh...15b3017S,cho14,2015MNRAS.453.3999M}. Whether or not this scale is reached depends on the age of the system, on the magnetic Reynolds number and on the nature of the turbulent forcing \citep[][]{fed14,bm16}.  In addition,  {\it astrophysical} sources at lower redshift ($z \leq 6$) inject magnetic fields into the cosmos. Galaxies can seed magnetic fields in large-scale structures, in an inside-out fashion: star formation drives winds of magnetized plasma into the circumgalactic medium \citep[e.g.][]{Kronberg..1999ApJ,Volk&Atoyan..ApJ.2000,donn09,2006MNRAS.370..319B,sam17}, and into voids \citep[][]{beck13}, while active galactic nuclei (AGN) can magnetize the central volume of clusters \citep[e.g.][]{2008A&A...482L..13D,xu09} and even the intergalactic medium during their quasar phase \citep[][]{2001ApJ...556..619F}. The efficiency of any magnetization process connected to galaxy formation is predicted to drop with the degree of overdensity (over the cosmic mean density), due to the dilution of injected fields with the expansion of the Universe as well as the fact that there are fewer sources in less dense regions. Some authors have suggested a lower limit to magnetic fields in voids of $\sim 10^{-15} ~\rm G$, due to the activity of magnetized outflows from isolated dwarf galaxies  \citep[][]{beck13}. A few studies obtained volume-averaged limits on magnetic fields on cosmological scales: using different methods, field strengths $\leq \rm~nG$ have been found \citep[][]{Blasi.Burles..1999, 2010PhRvD..82h3005K,2016A&A...596A..22B}. However, it is difficult to translate this into magnetic field strengths in cosmological objects such as galaxy clusters. 

Galaxy clusters host magnetic fields of several $\rm \mu G$ with coherence scales of several tens of $\rm kpc$, up to a few $\rm Mpc$ from the cluster center that seems to be consistent with various models for the magnetization of the Universe. 
A few cosmological magneto-hydrodynamical simulations could qualitatively reproduce the typical magnetization of galaxy clusters, starting from either fully {\it primordial} \citep[e.g.][]{do01,br05,bo11,ruszkowski11,va14mhd,2015MNRAS.453.3999M,2017arXiv170703396M} or {\it astrophysical} \citep[i.e. driven by AGN and/or star formation, e.g.][]{donn09,xu09,2012ApJ...759...40X} mechanisms.
However, in the periphery of galaxy clusters the magnetic field profiles resulting from different models is expected to diverge because dynamo amplification is less efficient as gas motions become increasingly supersonic \citep[e.g.][]{ry08, br11, 2013NJPh...15b3017S,va14mhd}.

However, observations of cosmological magnetic fields will remain complex and hard interpret, given the weakness of the expected signal and the combination with other effects. In this contribution, we will discuss how the combination of state-of-the-art MHD simulations in cosmology and  observations may be crucial to reveal the origin of extragalactic magnetic fields.  In Sec.~\ref{sim} we present our new suite of  numerical simulations.  In Sec.~\ref{res} we present our results, first by discussing the 
three-dimensional properties of magnetic fields (Sec.\ref{3d}), and second by predicting their signature in various observations: synchrotron radio emission (Sec.~\ref{radio}), Faraday Rotation (Sec.~\ref{rm}), Ultra-high Energy Cosmic Rays (Sec:~\ref{uhecrs}) and Fast Radio Bursts (Sec.~\ref{frb}). Our discussion and conclusions are given in Sec.~\ref{conclusion}, while technical details on the numerical methods are given in the Appendix.

\section{Cosmological magneto-hydrodynamical simulations}
\label{sim}

Simulating the evolution of extragalactic magnetic fields requires the joint modeling of the
evolution of dark matter (DM), gas and magnetic fields ab-initio in expanding space. In addition to gas compression, one of the key processes behind the emergence of large-scale fields within structures is the small-scale dynamo,  driven by turbulent motions \citep[e.g.][]{su06,ry08,po15}.  The simulation of the turbulent dynamo is hard because of the difficulty to reach the required magnetic Reynolds numbers of $R_m \gg 100$ for the exponential stage of the dynamo amplification to start \citep[e.g.][]{bm16}. 

The simulations presented here have been performed with a customized version of the {\enzo} code \citep{enzo14}.
{\enzo} solves the equations describing the dynamics of the two main matter components in the Universe, baryonic matter and DM, driven by the gravitational field generated by the combined mass distributions.
The DM component is evolved using a particle-mesh N-body method, while the baryonic component is represented as an ideal fluid, discretized on a mesh. The gravitational potential of the baryonic plus DM is calculated solving the Poisson equation through a combined multi-grid plus Fast Fourier Transform approach.

The magneto-hydrodynamical (MHD) module used in our simulations is the conservative Dedner formulation \citep[][]{ded02} which uses hyperbolic divergence cleaning to keep the $\nabla \cdot \vec{B}$ term as small as possible. The MHD solver adopts the PLM (Piecewise Linear Method) reconstruction, fluxes at cell interfaces are calculated using the using the Harten-Lax-van Leer (HLL) approximate Riemann solver. Time integration is performed using the total variation diminishing (TVD) second-order Runge-Kutta (RK) scheme \citep[][]{1988JCoPh..77..439S}. Recent papers have shown that this MHD approach is robust and competitive compared to other MHD solvers \citep[][]{kri11,2014ApJ...783L..20P,hop16}. In our version of the code we used the GPU-accelerated MHD version of {\enzo} by \citet[][]{wang10}. 

Although the use of AMR helps in resolving the turbulent flow inside forming structures \citep[e.g.][]{va17turb,wi17} in the simulations presented here we use a fixed resolution approach, using large ($\geq 1024^3$ ) grids.  This has the advantage of providing the best resolved view of magnetic fields even in low-density regions, which would  typically not be refined in AMR schemes. However, these regions are important if we look at the effects of cosmic magnetic fields on the propagation of ultra-high energy cosmic rays or Faraday Rotation. This choice allows high resolution in cluster outskirts and filaments, characterized by modest over-densities ($\rho/\langle \rho \rangle \sim 1-50$).  Dedicated studies of the amplification of magnetic fields in the innermost regions of galaxy clusters, based on AMR resimulations, will be presented elsewhere.

\begin{table}
\begin{center}
\caption{Main parameters of runs in the Chronos++ suite of simulations. For supermassive black hole and star particles, only the feedback parameters are given. For more information, see the Appendix.}
\footnotesize
\centering \tabcolsep 2pt
\begin{tabular}{c|c|c|c|c|c|c|c|c}
  $N_{\rm grid}$ & $\Delta x$ & $L_{\rm box}$  & cooling & BH feedback & SF feedback & $B_0$   & other sources of B & ID\\ 
   &   [kpc] & [Mpc] & & & & [nG] & & \\  \hline 
  $2400^3$ & 83.3 & 200  & n & n & n & 1 & n & P\\
 $2400^3$ & 83.3 &200  & primordial & $E_{\rm BH}=10^{58} \rm erg$& n & 0.01  & AGN ($\epsilon_B=0.01$) & CF1 \\
 $2400^3$ & 83.3 &200  & primordial & $E_{\rm BH}=5 \cdot 10^{59} \rm erg$& n & 0.01 & AGN ($\epsilon_B=0.01$) & CF2 \\ \hline 
      $1024^3$ & 83.3 & 85 & n & n & n &  1 & no &  P\\   
      $1024^3$ & 83.3 & 85 & n & n & n &  1 (Zeld.) & no &  Pz\\
      $1024^3$ & 83.3 & 85 & n & n & n &  1 (Zeld. 2nd) & no &  Pz2\\
      $1024^3$ & 83.3 & 85 & n & n & n &  $10^{-9}$  & dynamo, $\epsilon_{\rm dyn}(\mathcal{M})$  & DYN1\\
    $1024^3$ & 83.3 & 85 & n & n & n &  $10^{-9}$  & dynamo, $\epsilon_{\rm dyn}=0.02$  & DYN4\\
    $1024^3$ & 83.3 & 85 & n & n& n &  $10^{-9}$  & dynamo, $10 \cdot \epsilon_{\rm dyn}(\mathcal{M})$  & DYN5\\
    $1024^3$ & 83.3 & 85 & n & n & n &  $10^{-9}$  & dynamo, $\epsilon_{\rm dyn}=0.1$  & DYN6\\
    $1024^3$ & 83.3 & 85 & n & n & n &  $10^{-9}$  & dynamo, $\epsilon_{\rm dyn}(\mathcal{M})=0.04$  & DYN7\\
    $1024^3$ & 83.3 & 85 & n & n & n &  1  & dynamo, $\epsilon_{\rm dyn}(\mathcal{M})$  & DYN8\\
         $1024^3$ & 83.3 & 85 & primordial & n & n &  $1$  & n  & C\\
            $1024^3$ & 83.3 & 85 & primordial & $E_{\rm BH}=10^{58} \rm erg$ & n &  $10^{-9}$ & AGN($\epsilon_{b}=0.01$)  & CF1\\
            $1024^3$ & 83.3 & 85 & primordial & $E_{\rm BH}=5 \cdot 10^{59} \rm erg$ & n &  $10^{-9}$  & AGN($\epsilon_{b}=0.01$)  & CF2\\
        $1024^3$ & 83.3 & 85 & chemistry & n & $\epsilon_{SF}=10^{-8}$&  $10^{-9}$  & $\epsilon_{b,SF}=0.01$  & CSF1\\
           $1024^3$ & 83.3 & 85 & chemistry & n & $\epsilon_{SF}=10^{-7}$&  $10^{-9}$  & $\epsilon_{b,SF}=0.1$  & CSF2\\
              $1024^3$ & 83.3 & 85 & chemistry & n & $\epsilon_{SF}=10^{-6}$&  $10^{-9}$  & $\epsilon_{b,SF}=0.1$  & CSF3\\
                  $1024^3$ & 83.3 & 85 & chemistry & n & $\epsilon_{SF}=10^{-7}$& 1  & no & CSF4\\
              $1024^3$ & 83.3 & 85 & chemistry & n & $\epsilon_{SF}=10^{-8}$&  0.01  & $\epsilon_{b,SF}=0.1$  & CSF5\\
   $1024^3$ & 83.3 & 85 & chemistry & $\epsilon_{BH}=0.05$& $\epsilon_{SF}=10^{-8}$&  $10^{-9}$  & $\epsilon_{b,SF}=0.1$, $\epsilon_{b,BH}=0.01$  & CSFBH1\\
     $1024^3$ & 83.3 & 85 & chemistry & $\epsilon_{BH}=0.05$& $\epsilon_{SF}=10^{-8}$&  $10^{-9}$  & $\epsilon_{b,SF}=0.1$, $\epsilon_{b,BH}=0.01$  & CSFBH2\\
       $1024^3$ & 83.3 & 85 & chemistry & $\epsilon_{BH}=0.05$& $\epsilon_{SF}=10^{-6}$&  $10^{-9}$  & $\epsilon_{b,SF}=0.1$, $\epsilon_{b,BH}=0.1$  & CSFBH3\\
         $1024^3$ & 83.3 & 85 & chemistry & $\epsilon_{BH}=0.05$& $\epsilon_{SF}=10^{-6}$&  $10^{-9}$  & $\epsilon_{b,SF}=0.1$, $\epsilon_{b,BH}=0.1$  & CSFBH4\\
           $1024^3$ & 83.3 & 85 & chemistry & $\epsilon_{BH}=0.05$& $\epsilon_{SF}=10^{-6}$&  $10^{-9}$  & $\epsilon_{b,SF}=0.1$, $\epsilon_{b,BH}=0.1$  & CSFBH5\\
                $1024^3$ & 83.3 & 85 & chemistry & n & $\epsilon_{SF}=10^{-6}$&  1  & $\epsilon_{b,SF}=0.1$ , dynamo $\epsilon_{\rm dyn}(\mathcal{M})$ & CSFDYN1 \\

  \end{tabular}
  \end{center}
\label{table:tab1}
\end{table}

\subsection{The Chronos++ Suite of simulations}

We have produced a large suite of cosmological MHD simulations with {\enzo}, the "Chronos++ suite" {\footnote{http://cosmosimfrazza.myfreesites.net/the\_magnetic\_cosmic\_web}} in order to explore various scenarios for the emergence of  large-scale magnetic fields.
In the past, only few attempts have been made to compare the outcomes of competing scenarios for the origin of magnetic fields in galaxy clusters \citep[][]{donn09,2015MNRAS.453.3999M}. Most authors have focused on specific mechanisms for the origin of cosmic magnetic fields \citep[e.g.][]{do99,2008A&A...482L..13D,xu09,2017arXiv170703396M}. 
Our aim here is to produce the most complete survey of realistic models of magnetogenesis and model observations of large-scale fields. 
This suite includes three main classes of runs: {\it primordial}, {\it dynamo} and {\it astrophysical} magnetogenesis scenarios.  We give a brief explanation of each model below, while a more detailed explanation of the numerical aspects of each implementations is given in the Appendix:  

\begin{itemize}
\item{\it primordial} models: Here, we assume a volume-filling magnetic seed field at the beginning of the simulation  ($z_{\rm in}=38$). This field is then subject to compression, rarefaction and amplification in the course of structure formation.  The simplest way of initializing the seed field is by imposing a uniform value for every component of the initial magnetic field $B_0=B(z_{\rm in})$.  In addition, we generated seed magnetic fields with the same spatial structure of the matter perturbations using the standard Zeldovich approximation (see Sec.~\ref{zeld}). While the first approach mimics a scenario in which large-~cale ($\geq 100 ~\rm Mpc$) magnetic field domains are generated due to inflationary mechanisms, the second  mimics high-z seeding mechanisms that are related to  very early structure formation (e.g. Biermann battery, ionization fronts around first stars, etc). In order to compare to the outcome of the simpler uniform initialization, the r.m.s values of the field generated in the Zeldovich approximation are renormalized within the volume so that $\sqrt{|<B^2>|} = B_0$. In this project we compared to set of initial conditions obtained in this way, with two different spatial orientation of the initial magnetic field vectors. For the runs presented here, we  used a value of $B_0=1 ~\rm nG$ (comoving) which is closed to the maximum allowed by  the most recent constraints from the CMB \citep[][]{PLANCK2015}, i.e. $\sim 4 \rm ~nG$. In most of our simulations of primordial scenarios, we use a non-radiative setup for gas physics. This is particularly helpful  to understand the dependence of the magnetic fields on the spatial resolution and thus the effective Reynolds number \citep[e.g.][]{va14mhd}. It also helps to disentangle the role of gas cooling, feedback and dynamo amplification in the growth of simulated magnetic field (which is otherwise non-trivial, \citealt{2015MNRAS.453.3999M}). 
\item {\it dynamo} models: Here, we start with much weaker seed fields ($B_0 = 10^{-9} \rm nG$) but we allow for the dynamo amplification of fields by the solenoidal turbulence developed in large-scale structures. As the spatial resolution in our runs is often too coarse to allow for the development of a saturated dynamo, we implemented a simplistic  subgrid model for the dynamo, that computes the fraction of turbulent (solenoidal) kinetic energy flux which should go onto the amplification of seed fields \citep{po15,bm16}.  We rely on the  numerical survey of turbulence in a box runs by  \citet{fed14}, and use their fitting formulas to the observed growth of magnetic fields as a function of the local Mach number and plasma parameters. See  \ref{sg} for more details. 
\item {\it astrophysical} models: In these models magnetic fields are seeded by a) stellar activity and/or b) supermassive black holes at the center of active galactic nuclei. Star formation is modelled following the prescription by  \citet[][]{2003ApJ...590L...1K}, with tunable star formation efficiencies, timescales and feedback parameters.  We explored several combination of possible parameters, by benchmarking our models against the observed cosmic star formation rate (see \ref{sfr}).  We implemented the possibility of accounting for magnetic feedback energy from "star forming particles", with a tunable efficiency usually set to $\epsilon_{\rm SF,b}~\sim 1-10 \%$ of the total feedback energy. We refer to \ref{sfr} for more details.  The growth of supermassive black holes and their thermal/magnetic feedback on the surrounding gas is simulated with "BH particles", injected  at the location of massive halos at $z=4$ \citep[][]{2011ApJ...738...54K}. The feedback energy is released as an extra thermal energy output from each black hole particle,  and also in this case we link each episode of AGN feedback to the release of magnetic field from magnetized dipoles jets (assuming efficiencies in the range $\epsilon_{\rm BH,b}=1-10 \%$). In our largest $2400^3$ runs we also used an alternative approach where impulsive feedback from high-density regions ($\geq 10^{-2} ~\rm part/cm^3 $) undergoing cooling is released with a fixed thermal energy per event, $E_{\rm BH}=10^{58}-5 \cdot 10^{59}$ erg. This approach yields a fairly good representation of the large-scale impact of the combined sellar and AGN feedback in simulations too expensive to allow also for the inclusion of more chemistry, stellar or BH particles  (see \ref{bh} for more details). 
\end{itemize}

In all runs we assumed a $\Lambda$CDM cosmological model, with density parameters $\Omega_{\rm BM} = 0.0478$, $\Omega_{\rm DM} =
0.2602$,  $\Omega_{\Lambda} = 0.692$, and a Hubble constant $H_0 = 67.8$ km/sec/Mpc \citep[][]{2016A&A...594A..13P}.  All runs  started at $z=38$ and  have the constant spatial  resolution of  $83.3 ~\rm kpc/cell$ (comoving) and the constant mass resolution for DM of  $m_{\rm dm}=6.19 \cdot 10^{7}M_{\odot}$ per particle.  The most relevant parameters that controls the magnetization levels in our runs are given in Tabl.~1, while a more detailed list of parameters involved in the simulation of star forming particles and supermassive black hole is given in the Appendix (Tab.~2). 

In a  first set of runs we resimulated a  (85 Mpc)$^3$ volume with $1024^3$ cells and DM particles with many physical variations, in order to  monitor the effect of gas physics and magnetic field scenarios on the large-scale distribution of magnetic fields. Manifestly, the resolution in our runs is too coarse to properly resolve the complex and multi-phase process of star formation in galaxies, for which many more advanced models have been developed over the years \citep[e.g.][just to least a few seminal works on the subject]{2012MNRAS.423.1726S,2013MNRAS.429.3068T,  2014MNRAS.445..175G,2014ApJS..210...14K,2015MNRAS.446..521S, 2016MNRAS.463.3948D,2017Galax...5...35D}. Our main goal here is to monitor the impact of {\it effective} (i.e. resolution dependent and ad-hoc calibrated) galaxy formation models on the large-scale generation of magnetic fields. 

In this work we focus on a subset of the 23 most realistic runs in the Chronos++ suite. Besides gas and DM, the radiative runs of this dataset also followed the evolution of  seven chemical species (HI, HII, HeI, HeII, HeIII, free electrons and "metals").  While in most runs we only focus on one set of models, in a couple we combine them, to see the maximum concurrent effect of all sources of magnetization (e.g. runs DYN8 and CSFDYN1). 
The typical  time needed to complete these runs from $z=38$ to $z=0$ is $\sim 100,000$ core hours for each non-radiative run and $\sim 400,000$ core hours (on average) for our runs with cooling, chemistry, star formation and BH particles. 

A second set of three larger simulations of a (200 Mpc)$^3$ box was run using $2400^3$ cells/DM particles to allow us a better comparison with the large volumes which radio (e.g. LOFAR, MWA, SKA etc ) and X-ray (e.g. eRosita, Athena) observations will probe.  In this second case, we simulated a simple primordial scenario ($B_0=1 ~\rm nG$)  and two variations of the astrophysical scenarios (with a fixed energy release per feedback event of $E_{\rm BH}=10^{58} ~\rm ergs$ or of   $E_{\rm BH}=5 \cdot 10^{59} ~\rm ergs$, in both cases assuming $\epsilon_{\rm BH,b}=1 \%$).  In the astrophysical scenarios, we also included radiative equilibrium gas cooling, for a constant metallicity of $Z=0.3 ~Z_{\odot}$, but no other chemistry in order to limit the memory requirement of the simulation. 
 The typical computing time needed to complete these runs from $z=38$ to $z=0$ is $\sim 1,5$ million core hours for the non-radiative run and $\sim 4,5$ million hours  for the radiative runs.

The most computationally demanding simulations have been run on the Piz Daint HPC system, operated by the Swiss National Supercomputing Centre (ETHZ-CSCS{\footnote{www.cscs.ch}}), a Cray XC50/XC40 supercomputer with more than 5000 computing nodes, each equipped with 12 cores Intel Xeon E5-2690/2695 CPUs and an NVIDIA Tesla P100 GPUs. Our $1024^3$ simulations  typically used 64 computing nodes of Piz-Daint, while our $2400^4$ simulations required up to 2400 computing nodes. A subset of our $1024^3~$ simulations was run on the Jureca cluster of the Juelich Supercomputing Centre (JSC), using 64 standard nodes.

\begin{figure}
\includegraphics[width=1.1\textwidth]{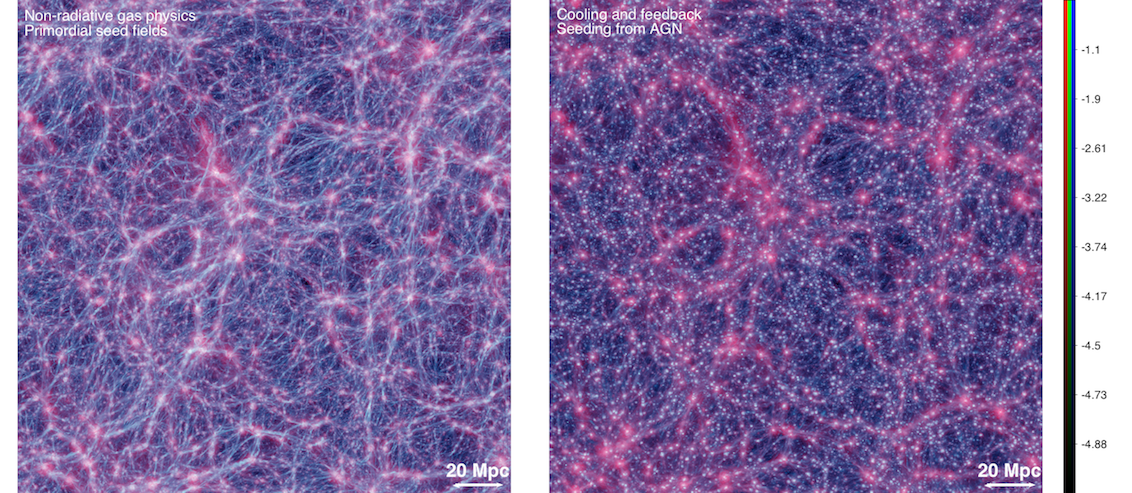}
\caption{Volume rendering of the average temperature  (red colors) and average  magnetic field strength (green+blue)  along the line of sight for our 
largest $2400^3$ run, for the non-radiative run with a primordial scenario ($B_0=1 ~\rm nG$ for the origin of magnetic fields (left) or a cooling \& feedback run where magnetic fields have been injected by AGN (right). For the both quantities we show the mass-weighted average along the line of sight. In both panels the temperature ranges from $10^{4} \rm K$ to $10^{8} \rm K$ and the magnetic field strength from $10^{-4}\rm \mu G$  to $1 ~\rm \mu G$  (with a $\log_{\rm 10}$ stretching of colors, see colorbar ).}
\label{fig:map}
\end{figure}

\begin{figure}
\includegraphics[width=1.1\textwidth]{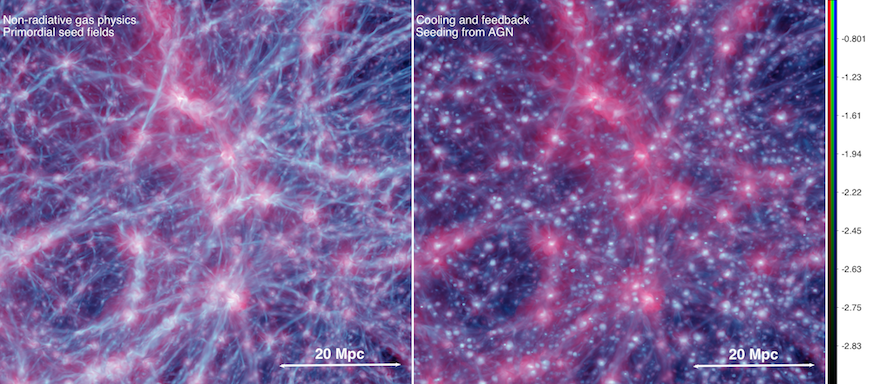}
\caption{Close up view for the central region of Fig.\ref{fig:map} (same meaning of colors).} 
\label{fig:map2}
\end{figure}

Figs.~\ref{fig:map}-\ref{fig:map2} show the projected gas temperature and  magnetic field strength for the two extreme scenarios explored in our our largest $2400^3$ runs  at $z=0$: the  primordial (P) and the astrophysical scenario (run CF1), for two different levels of zoom. Clearly, the two models lead to a very similar magnetic fields inside the densest knots of the cosmic web such as clusters or groups of galaxies, but they increasingly diverge in cluster outskirts, filaments and voids. The large-scale morphology of the magnetic cosmic web is different in the two scenarios:  it resembles the  matter cosmic web in the primordial scenario, but is mostly made of percolating magnetic "bubbles" if the sources of magnetic fields are located in galaxies.

\section{Results}
\label{res}

\begin{figure}
\includegraphics[width=1.1\textwidth]{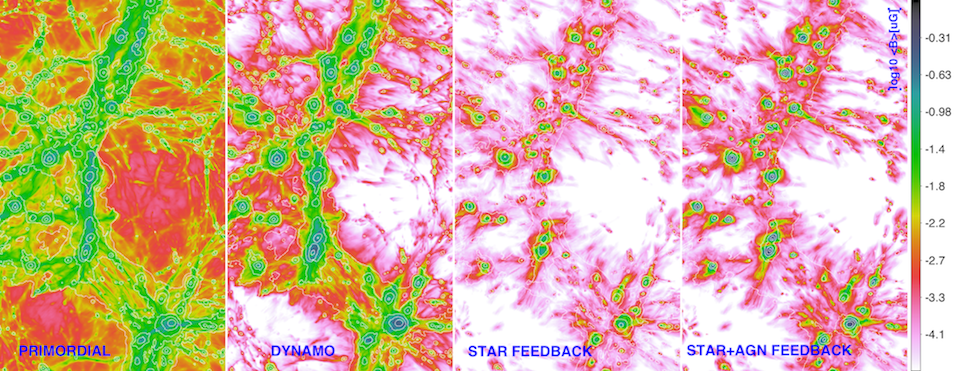}
\caption{Projected mean  (mass-weighted) magnetic for four runs of the Chronos++ suite at $z=0$: a) primordial seeding model (P); b) dynamo model (DYN7);  c) model with  injection of magnetic fields by star-forming particles (CSF1); d) model including also the magnetic field injection from supermassive black holes (CSFBH4). The additional contours display the projected pressure for the same volume. The side of the image is $35 \times 70$ Mpc. }
\label{fig:map0}
\end{figure}

\subsection{The distribution of magnetic fields across cosmic environment}
\label{3d}

Fig.~\ref{fig:map0} shows the mass-weighted magnetic fields at $z=0$, as simulated in four representative resimulations of the (85 Mpc)$^3$ box. Evidently, the distribution of gas pressure is nearly identical in all runs, at least on $\geq ~\rm Mpc$ scales where cooling and feedback effects play a minor role. However, the three-dimensional distributions of the magnetic fields in the different scenarios show significant differences. 
Inside clusters and groups of galaxies, the magnetic fields are similar in all cases ($\sim 0.1-1 ~ \rm \mu G$), which is not surprising because all these recipes have been designed to match the magnetic field typically observed in large-scale structures. However, the agreement gets much worse in the periphery of large-scales structures because of the different efficiency with which each model produces large-scale fields. In the primordial scenario, voids are magnetized at the $\sim 0.1 ~\rm nG$ level, which is the remnant of the initial seed field, rarefied by the density drop $\propto n^{2/3}$. Filaments and clusters outskirts have a magnetic field $\sim 10-100$ times larger than in voids, due to compression and amplification. 

On the other hand, if the observed magnetization in large-scale structures follows from the dynamo amplification of weaker initial fields (second panel), the magnetic fields in voids will be orders of magnitude smaller at z=0 ($\sim 10^{-16} ~\rm G$). The magnetic fields caused by dynamo amplification in filaments and cluster outskirts are roughly as large as in the previous scenario ($\sim 10-100 \rm ~nG$), but with a strong dependence on gas density because of the increased level of solenoidal turbulent motions in denser regions \citep[e.g.][]{ry08, va14mhd}.

In the other astrophysical scenarios, filaments are much less magnetized, typically at the $\sim ~\rm nG$ level if the initial seed field is negligible. The distribution of magnetic fields in the cosmic web mirrors that of active galaxies, and magnetization proceeds from "blobs" expanding from the densest objects and these blobs eventually merge. A few $\sim \rm ~Mpc$ away from supermassive black holes and their induced AGN activity, we do not see much difference between models in which the magnetization only comes from star-powered outflows (third vs fourth panel).

\begin{figure}
\includegraphics[width=0.33\textwidth]{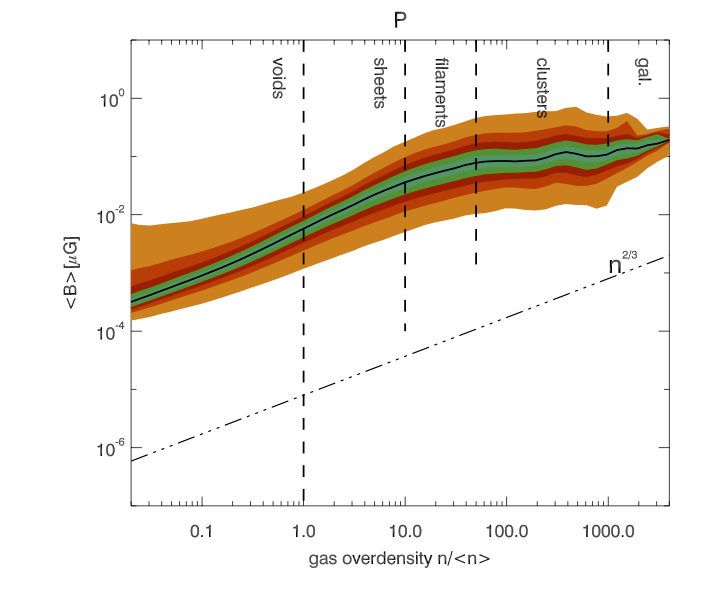}
\includegraphics[width=0.33\textwidth]{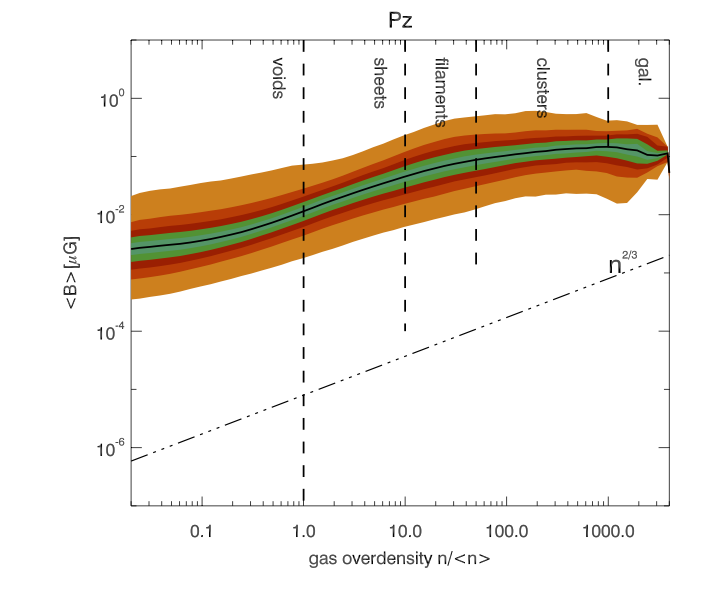}
\includegraphics[width=0.33\textwidth]{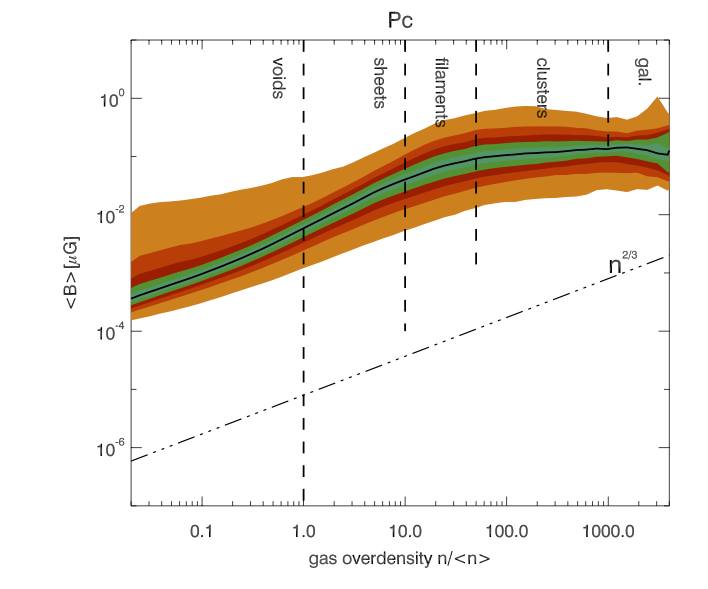}
\includegraphics[width=0.33\textwidth]{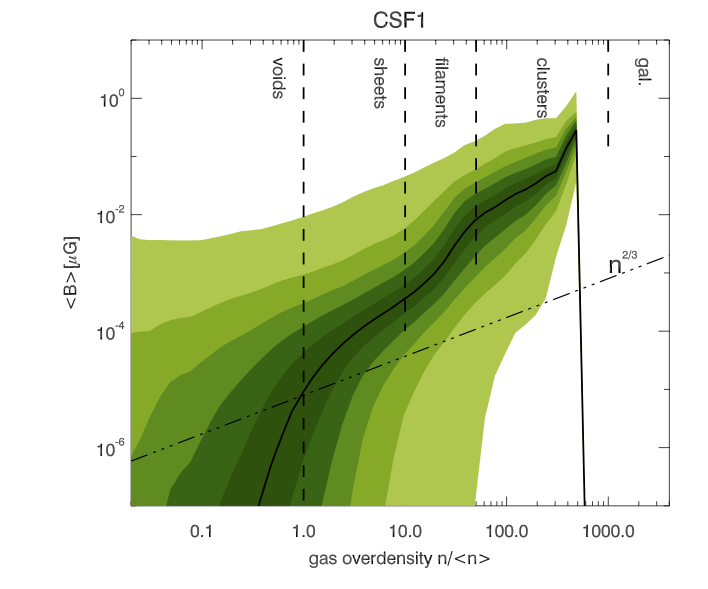}
\includegraphics[width=0.33\textwidth]{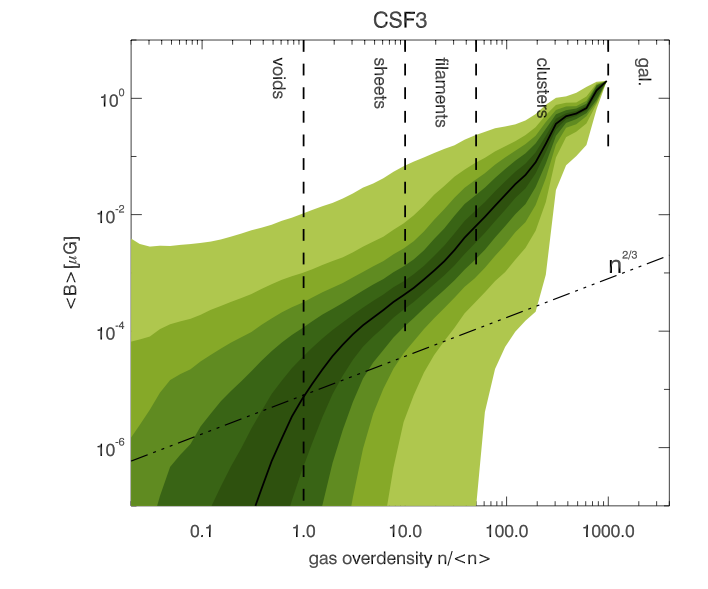}
\includegraphics[width=0.33\textwidth]{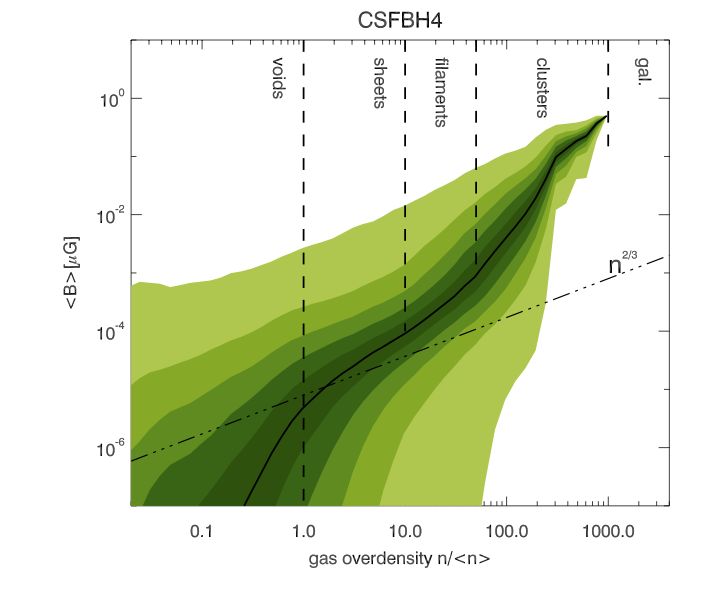}
\includegraphics[width=0.33\textwidth]{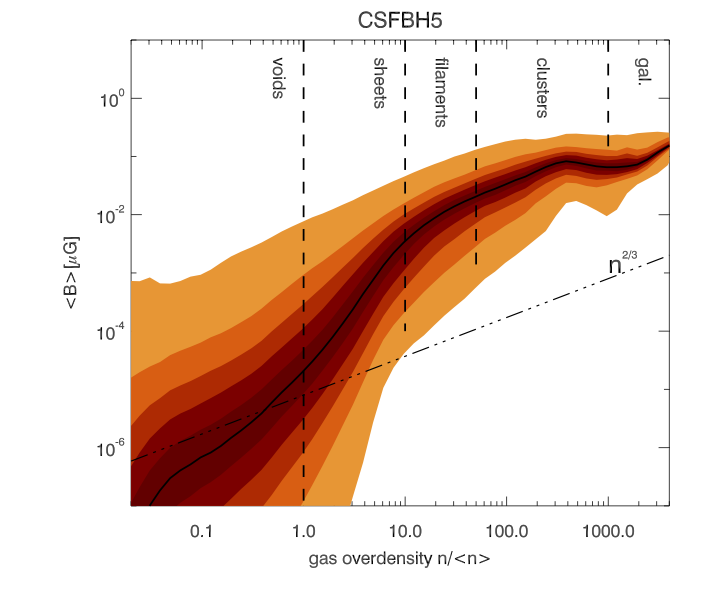}
\includegraphics[width=0.33\textwidth]{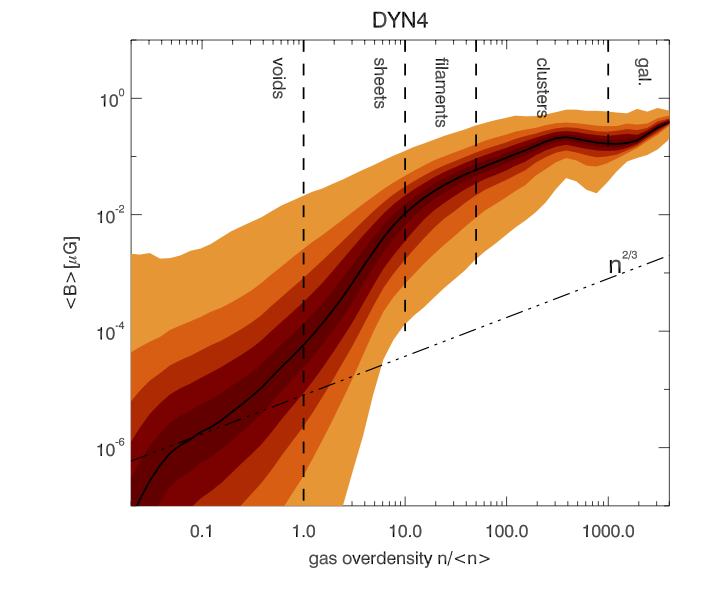}
\includegraphics[width=0.33\textwidth]{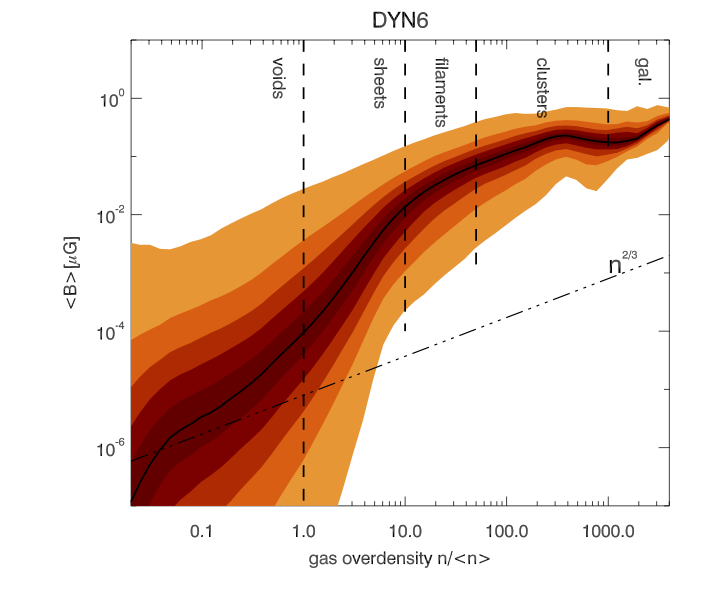}
\includegraphics[width=0.33\textwidth]{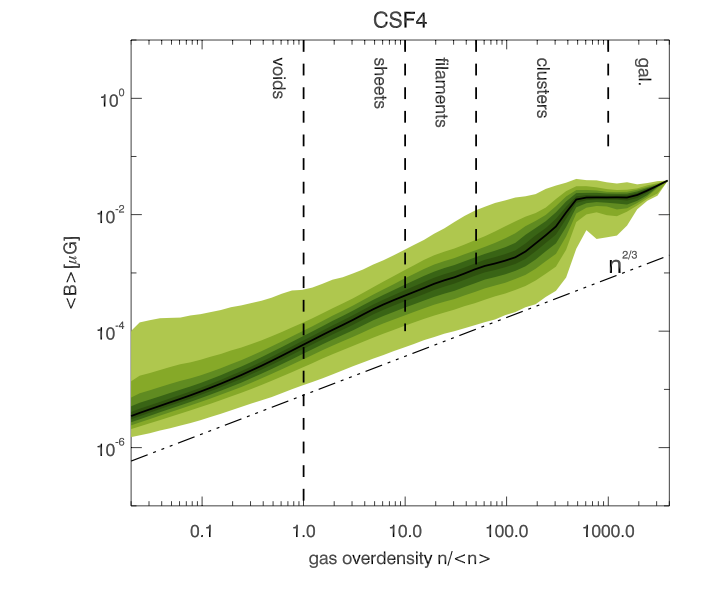}
\includegraphics[width=0.33\textwidth]{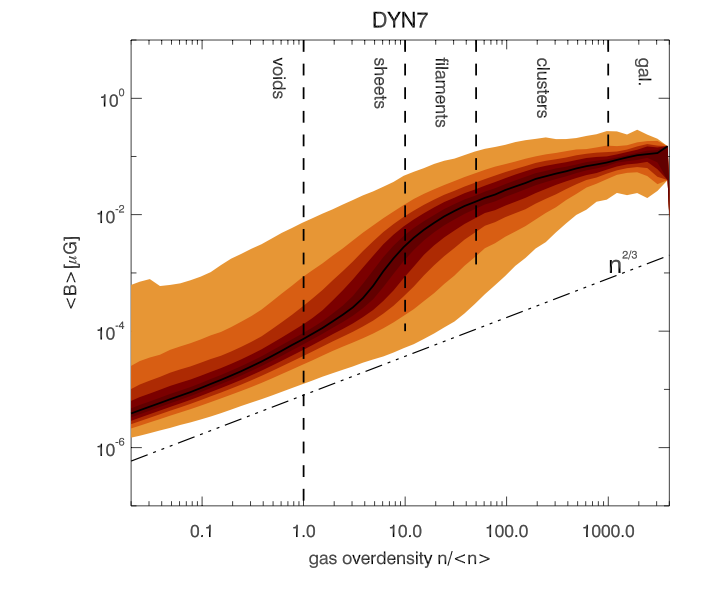}
\includegraphics[width=0.33\textwidth]{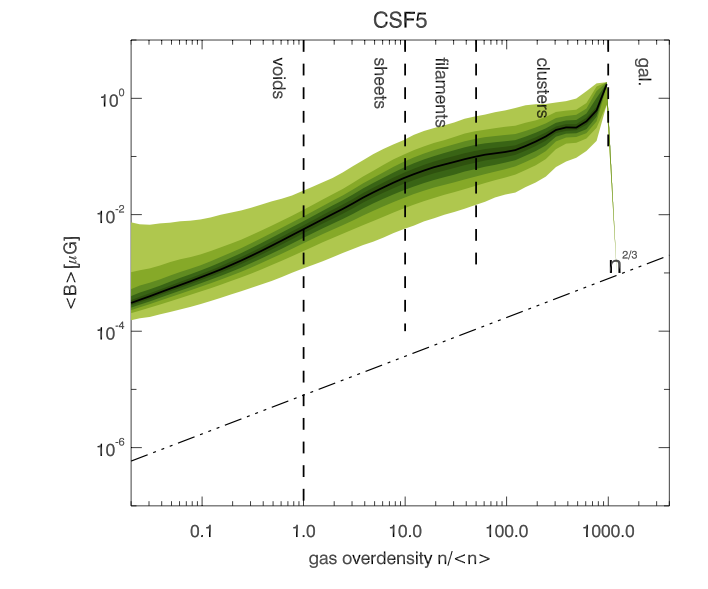}
\caption{Phase diagram ($B$,$n/\langle n \rangle)$ for 12 representative runs of the Chronos++ suite. The vertical lines marks the approximate overdensity regime of typical cosmic structures, the diagonal line shows the $B \propto n^{2/3}$ scaling. The various levels show the following percentiles: 1, 10, 20, 30, 40, 50,60, 70, 80, 90 and 99.}
\label{fig:phase1}
\end{figure}

In Fig.~\ref{fig:phase1},  we show the  ($B,\rho/\langle \rho \rangle$)  phase diagrams of twelve representative runs at $z=0$ (three primordial, astrophysical and dynamo run, plus three "mixed" models). The main points to take away from them are:

\begin{itemize}

\item only a very mild amplification with respect to the $B \propto n^{2/3}$ is observed in our primordial runs, that does not resolve the small-scale dynamo inside virialized halos {\footnote{With a tailored campaign of adaptive mesh refinement simulations of galaxy clusters, we find indeed that a achieving a  resolution of the order of $4 ~\rm kpc$, in a nearly uniform way inside the virial volume of halos is necessary to resolve the small-scale dynamo (Vazza et al., submitted). Hence we do not expect to recover realistic fields inside galaxy clusters because the dynamo  is artificially quenched or delayed by the lack of resolution \citep[e.g.][]{bm16}. }}. Moreover, the onset of a cooling flow compresses the field to even larger values at high densities. {\it [run P vs Pz vs Pc, top row.]}
\item Differences in the initial topology of the primordial field have a small effect on the overall  ($B,\rho/\langle \rho \rangle$)  relation. However, we observe larger fluctuations of magnetic fields in voids in the Zeldovich initialization. {\it[run P vs Pz]}
\item The effect of cooling, star formation and feedback from supermassive black holes significantly increases the scatter in the magnetic fields at all overdensities. This is an effect of outflows and intermittent activity from galaxy formation processes. {\it[second  vs first row]}
\item The maximum magnetic fields that can be achieved from star formation slightly depends on the assumed minimum density to trigger such process, and almost linearly scales with the magnetization efficiency from the stellar feedback. This is also consistent with the run-time magnetic output of the star formation recipes, see \ref{sfr}. {\it [CSF1 vs CSF3]} 
\item The additional seeding from supermassive black holes only has a modest impact on the global magnetisation. It is also limited to high-density regions if star formation is already included, mostly because the reservoir of cool gas which triggers feedback has been already expelled by star formation before large enough supermassive black holes can form. {\it [CSFBH4 vs CSF1 and CSF3]}
\item If we include the effect of dynamo amplification in a subgrid fashion, $\sim 0.1-1 \rm ~\mu G$ magnetization level of halos can be recovered even by starting with extremely weak fields.  The final magnetic field level scales almost linearly with the assumed magnetization efficiency of the dynamo. {\it [third row]}
\item For $B_0 > 0.1 \rm ~ nG$ seed fields, the effect of dynamo amplification is more important than compression in filament or in denser environments. For lower seed fields, the effect of dynamo amplification may be larger even at the density of cosmic sheets.   {\it[DYN7 vs DYN4 and DYN5]}. 
\item A scenario in which all sources of magnetization have maximum efficiency is not very dissimilar from a primordial model starting from $B_0=1 \rm~ nG$. However, the scatter in magnetization as a function of overdensity is smaller than in a purely astrophysical scenario because large primordial fields and the small-scale dynamo produce a floor of magnetization as a function of overdensity. {\it [CSF5 vs CSF1  and CSF3]} 
\end{itemize}

As evident from Fig.~\ref{fig:map0}, the median magnetic field in different scenarios clearly evolve differently in filaments or even lower-density structures. This is one of the most important properties we will use in the following sections to identify future observations. 
The amount of scatter in magnetic field strength measured at each overdensity is an additional key feature that discriminates between models: for example at the overdensity of filaments, we expect a roughly $\sim 100$ times larger scatter in magnetic fields in astrophysical models compared to primordial and dynamo models because density is not the only variable leading to a given magnetization level (e.g. also the past star-formation history contributes to it).

\begin{figure}
\includegraphics[width=0.95\textwidth]{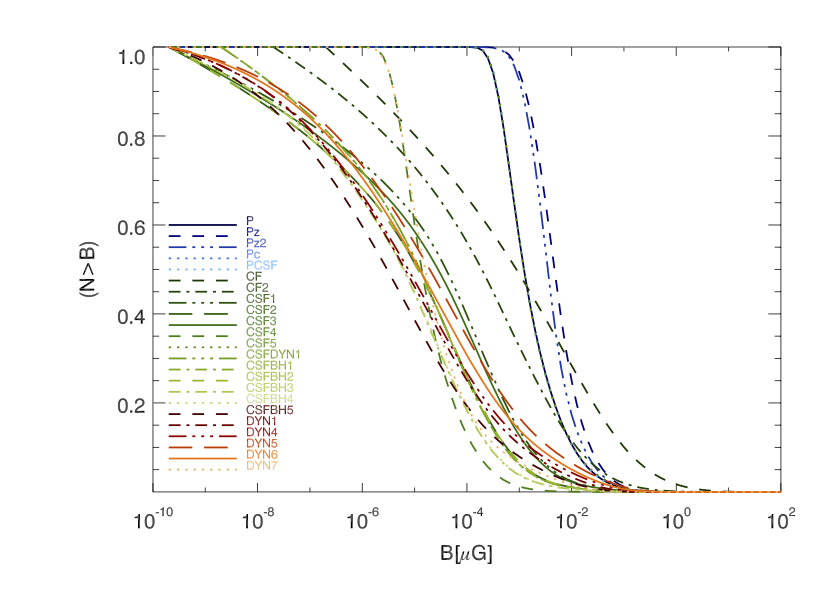}
\caption{Cumulative volume filling factor of magnetic fields in all runs in Tab.~1.}
\label{fig:pdf}
\end{figure}

The volume distribution function of all simulations is shown in Fig.~\ref{fig:pdf}. This statistics is particular important for processes that are mostly affected by the magnetization on the largest scales, such as the propagation of Ultra-High Energy Cosmic Rays \citep[][]{2014JCAP...11..031A,hack16}, as we will discuss in Sec.~\ref{uhecrs}. Roughly speaking, $\sim 50\%$ of the simulated volume in our primordial models has $\geq 1 ~\rm nG$, while the same is true only in  $\sim 10-15 \%$ of the simulated volume for dynamo models, and $\leq  10 \%$ for most astrophysical models.  

\begin{figure}
\includegraphics[width=0.95\textwidth]{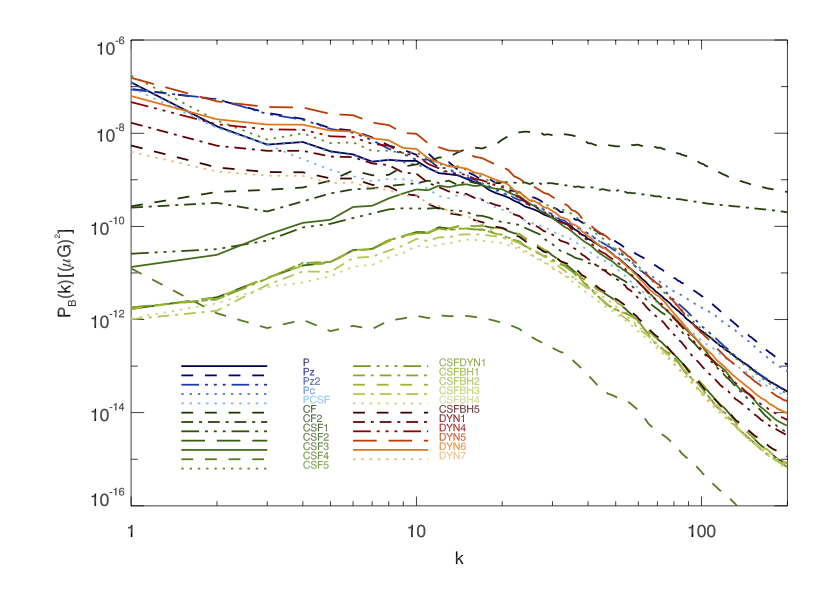}
\caption{Power spectra of magnetic field energy for  all runs in Tab.~1. Based on the 400$^3$ cells subvolume used to compute the spectra, $k=1$ here corresponds to $33.2 ~\rm Mpc$ and $k=200$ to $166 \rm ~ kpc$.  }
\label{fig:power}
\end{figure}

The large-scale topological distribution of simulated fields is captured by the three-dimensional power spectra of the magnetic field (here only the $x$-component is shown),  calculated with a standard Fast Fourier Transform (FFT) approach and assuming periodic boundary conditions (Fig.~\ref{fig:power}). 
Besides the obvious differences in normalization, it is evident that the astrophysical models concentrate more spectral power on scales of $\leq 2-3 ~\rm Mpc$, while most of magnetic energy in primordial and dynamo models comes from the largest scales ($\geq 10 ~\rm Mpc$).
While the difference in magnetic energy is $\geq 5$ orders of magnitude at low spatial wave-numbers, the difference is reduced to $\leq 1-2$ orders of magnitude at the highest wave-numbers. 
Furthermore, in most scenarios the typical slope of the magnetic energy spectrum is $\alpha_{\rm B} \approx -3.5 \div -4.5$ with $P_{\rm B} \propto k^{\alpha_{\rm B}}$. This is not far from the typical slope of the velocity spectrum on the same scales \citep[e.g.][]{ry08}, which in turn reflects the clustering of matter on these scales \citep[e.g.][]{2009MNRAS.397.1275W}.

\begin{figure}
\includegraphics[width=0.95\textwidth]{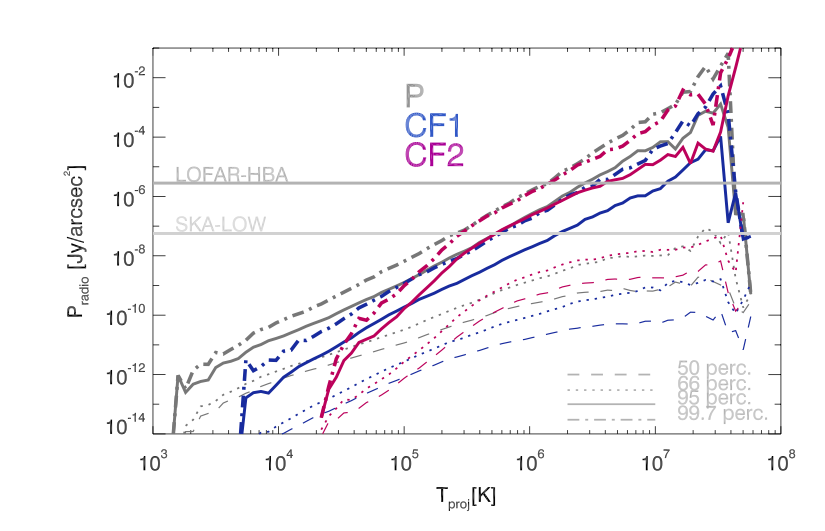}
\caption{Synchrotron emission at $\nu=120 ~\rm MHz$ as a function of projected gas temperature in the primordial and astrophysical seeding scenarios of our simulated (200 Mpc)$^3$ volume. Each linestyle shows the 50, 66, 95 and 99.7 percentile in the distribution.  The additional horizontal lines show the estimated sensitivity of surveys with LOFAR-HBA and SKA-LOW. } 
\label{fig:radio_temp}
\end{figure}

\subsection{Synchrotron radio emission from shock-accelerated electrons (in cluster outskirts and filaments)}
\label{radio}

Although the  warm-hot intergalactic medium (WHIM) should contain about a half of the total matter in the Universe \citep[e.g.][for a recent review]{2016xnnd.confE..27N}, its direct observation is limited to a few cases in  X-rays \citep[][]{2015Natur.528..105E,2017arXiv170405843A} or through the  Sunyaev-Zeldovich effect \citep[][]{2013A&A...550A.134P}. However,  the opening of the radio window on the magnetized cosmic web expected with the next generation of radio telescopes (LOFAR, MWA, MEERKAT and SKA) will change this. 

Considering that in most magnetogenesis models the memory of the initial seed field is preserved in filaments, the detection of the WHIM at radio wave lengths holds large promises.  The geometric and thermodynamical properties of the gas in filaments can be predicted across a wide range of scales in our simulations \citep{gh15}, and they show well-defined scaling relations which mirror the scaling relations of the DM within them \citep[e.g.][]{2014MNRAS.441.2923C}. Simulated filaments show very little evolution of their properties for $z \leq 1$ and their thermodynamical properties display strong correlations with the number and stellar mass of the galaxies they host \citep[][]{2016MNRAS.462..448G}. These correlations are confirmed by observations \citep[][]{alp14a}. 

Filaments and cluster outskirts should be surrounded by strong accretion shocks ($\mathcal{M} \gg	 10$), where cosmic rays may be accelerated \citep[e.g.][]{ry03}. Relativistic electrons accelerated at these shocks must emit polarized radio emission, analogous to the radio relics detected in merging clusters  \citep[e.g.][]{hb07,vw10,fe12,2012MNRAS.426...40B,fdg14}. This class of large-scale radio emission  has different properties from the other important class of diffuse emission in clusters, i.e. "radio halos" \citep[e.g.][]{fe08,fe12}, whose origin is most likely explained by the  Fermi II re-acceleration of mildly relativistic particles in the turbulent intracluster medium \citep[e.g.][]{2001MNRAS.320..365B,cassano05}. 
Despite of a few claims of possible detections \citep{2010A&A...511L...5G, 2013ApJ...779..189F},  the observation of large-scale shocks along filaments remains a challenge \citep[e.g.][]{2011JApA...32..577B}. 

In recent work, we have looked at the potential of existing and future surveys to detect the cosmic web in emission at radio wavelengths. We assumed particle acceleration by diffusive shock
acceleration {\footnote{Even though this model reproduces some properties of radio relics \citep[e.g.][]{sk11,2017MNRAS.470..240N}, it is challenged by several relics apparently associated with weak shocks \citep[e.g.][]{ka12,2013MNRAS.435.1061P,va14relics,va15relics}.}}, following the formalism by \citet{hb07}, and combined the Mach number and the magnetic fields measured in the simulation to predict the radio flux and the spectrum of synchrotron radiation. In the typical $B \leq 3.2 \mu G (1+z)^2$ regime of the WHIM and for strong shocks, the emission as a function of frequency ($\nu$) scales as 
\begin{equation}
P_{\nu} \propto  n_u \xi(\mathcal{M}) \cdot \mathcal{M}^3 c_s^3 S \cdot  B^2, 
\label{hb}
\end{equation}
where $n_u$ is the upstream gas density, $\xi(\mathcal{M})$ is the (unknown) acceleration efficiency of electrons, $c_s$ is the up-stream sound speed and $S$ is the shock surface. The dependence between the acceleration efficiency of electrons and the Mach number is expected to be a steep function of $\mathcal{M}$ for weak shocks rapidly saturate for $\mathcal{M} \geq 10$. In our simulations we assume $\xi_e \approx 7 \cdot 10^{-4}$ as a limiting value. 

We showed that the shocks inside the cosmic web may be detected by the next generation of radio telescopes  \citep[][]{va15ska,va15radio,va16radio}. The expected flux density is quite weak ($\leq \rm \mu Jy/arcsec^2$ at $\sim 100 ~\rm MHz$), typically diffuse on large angular scales ($\geq 1^\circ$) and highly polarized ($\sim 70\%$) {\footnote{It shall be noticed that the additional contribution from radio emitting secondary electrons produced in hadronic collisions \citep[e.g.][]{mi01} and/or dark matter annihilation \citep[e.g.][]{2017ApJ...839...33S} may add to the signal from shocked electrons. However, these two additional mechanisms are predicted to rapidly fall off outside of the innermost core of halos, and can be safely neglected in predicting the radio signal from the shocked cosmic web (see e.g. Appendix in \citealt{va15radio}).}}. The SKA-LOW and its low-frequency  precursors and pathfinders (e.g. LOFAR, MWA) may detect the tip of the iceberg of this emission, and allow to constrain the
$\xi_e \cdot B^2$ combination provided we know something about the thermal properties of the shock. 

With the same formalism, we predicted the radio flux for the different models in the Chronos++ suite. Here we focus on the large (200 Mpc)$^3$ resimulations, which gives us the largest sample of galaxy clusters and the closest match to the large radio survey of the future. 

\begin{figure}
\includegraphics[width=0.99\textwidth]{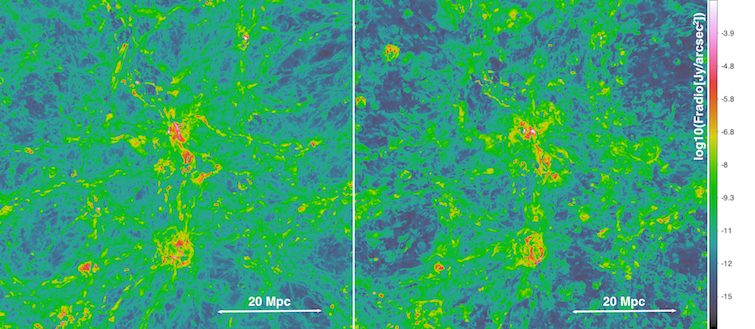}
\includegraphics[width=0.99\textwidth]{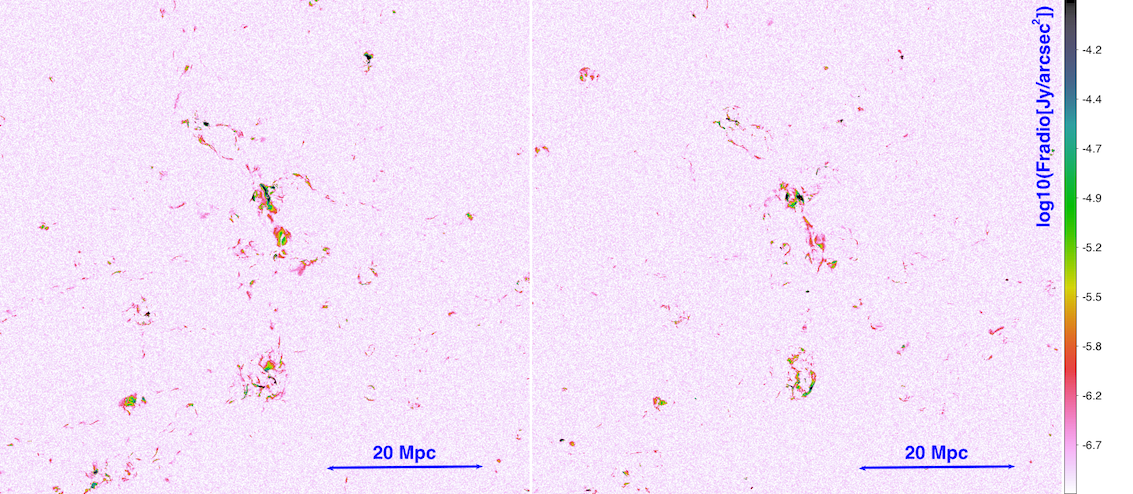}

\caption{Top panels: projected synchrotron radio emission from the same region of Fig.\ref{fig:map2} for our primordial (left) and AGN seeding scenario (CF1, right) for our  $2400^3$ runs at $z=0$. The line of sight considered here is $200 ~\rm Mpc$ and the closest structures are at a luminosity distance $d_L=100 ~\rm Mpc$ ($z \approx 0.024$). Bottom panels: for the same regions, we show the mock observation with SKA-LOW (assuming a central frequency of $\nu=260$ MHz, a resolution beam of $7.3"$, a maximum baseline of  $40 ~\rm km$ and a minimum baseline of $5 ~\rm m$. }
\label{fig:map_radio}
\end{figure}

Fig.~\ref{fig:map_radio} shows the predicted radio flux density at $z=0.024$ (corresponding to a luminosity distance $\approx 100 ~\rm Mpc$) and at the central frequency of $\nu =260\rm~MHz$, both in the primordial and in the AGN (strong) seeding scenario, for the same selection of Fig.~\ref{fig:map2}. While the radio~flux from within  clusters is remarkably similar, at the level of $\sim$ mJy/arcsec$^2$, such emission drops faster with decreasing overdensity in the AGN seeding scenario, declining by 1-2 orders of magnitude outside filaments. 

By correlating the radio emission with the mean mass-weighted temperature of our sky model, we can predict the typical relation between radio emission and gas properties across environments. In Fig.~\ref{fig:radio_temp},
we show the prediction for ($P_{\rm radio},T_{\rm proj}$) in our three runs of the (200 Mpc)$^3$ volume. Different percentiles of the distribution at each temperature bin are shown, to highlight the  fact that while the median emission is clearly different between scenarios, the high percentiles have similar values and are therefore more difficult to distinguish. Only the top $\sim 5 \%$ of the distribution is typically above the detection level for existing (e.g. LOFAR HBA) or future (e.g. SKA-LOW) radio surveys at low frequencies, in this case $\nu=$ 120 MHz. 

\begin{figure}
\includegraphics[width=0.49\textwidth]{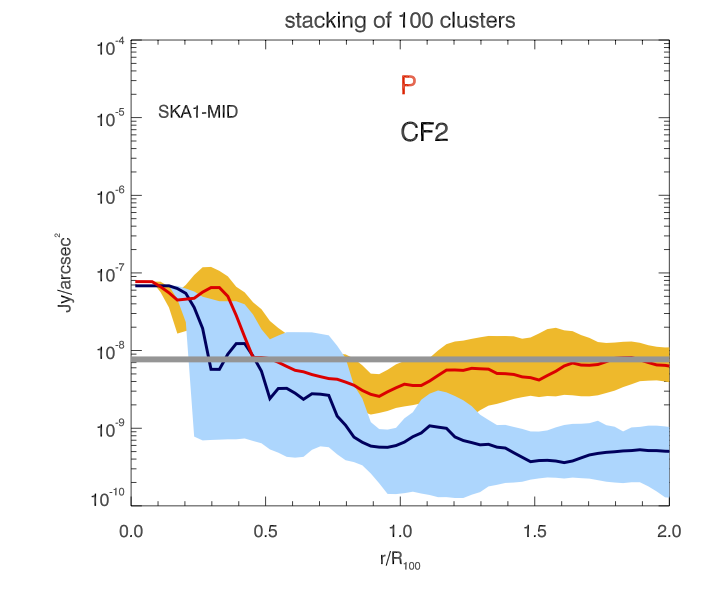}
\includegraphics[width=0.49\textwidth]{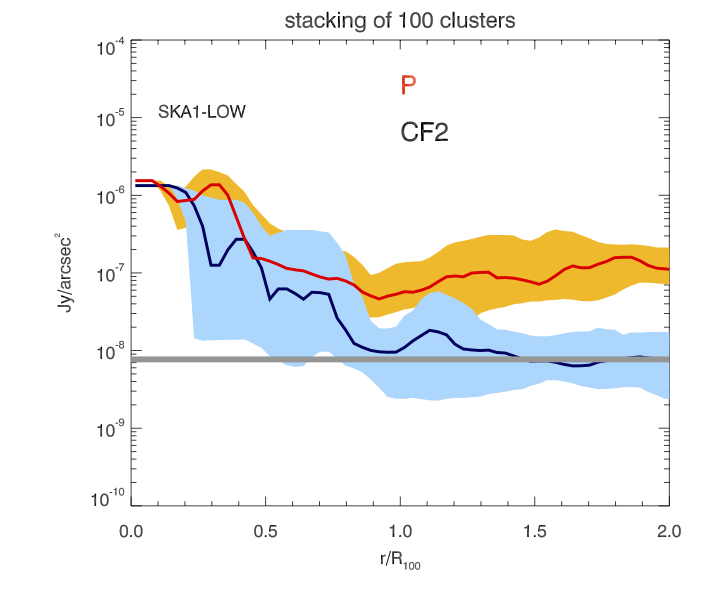}
\caption{Average ($\pm 1 \sigma$) radio emission profiles of 100 stacked simulated clusters ($M_{\rm 100} \geq 10^{14} M_{\odot}$) in the primordial and in the AGN (high power) resimulations of the $200^3 \rm Mpc^3$ volume at $z=0$. The left panel shows the stacking of mock radio maps observed at $1.4$ GHz with SKA-MID, while the right panel shows the stacking for SKA-LOW observations at $110$ MHz The horizontal gray lines show the approximated level of the thermal+confusion noise in both observational setups.}
\label{fig:radio_prof}
\end{figure}

This is evident also by looking at the lower panels of Fig.~\ref{fig:map_radio}, where we show the mock observation of the same region, assuming a $\sim 6$ hours integration for a survey with SKA-LOW survey at the central frequency of $\nu=260 ~\rm MHz$.  Here we followed the same procedure of our previous works (but with the most updated performances expected for SKA-LOW), including  the effect of thermal and confusion noise ($\sim 4.8 ~\rm \mu$Jy/beam) in the final image, the finite resolution beam ($\approx 7.3"$) and the minimum baseline of the observation ($5$ m).  The differences in the two models are now mostly obscured by the noise. Still, significant differences are  present, but they are best highlighted with the use of statistical techniques.
For example, in Fig.~\ref{fig:radio_prof} we show the result of the stacking of 100 mock radio maps of 
clusters with $M_{\rm 100} \geq 10^{14} ~ M_{\odot}$, in the primordial and in the AGN (high power) case at $z=0$.
The analysis has been performed both at  $\nu=1.4$ GHz for SKA-MID observations and at $\nu=120 ~\rm MHz$  for SKA-LOW observations, for which the approximated noise level (also including the estimate confusion noise level) has been overplotted as horizontal lines.  
While the two models display a similar average level of stacked flux in the innermost cluster regions, at both frequencies, the expected result of the stacking vary from $\sim 0.5 ~ R_{\rm 100}$, becoming statistically significant ($\geq 3 ~\sigma$) at $R_{\rm 100}$. However, only SKA-LOW should have the sensitivity on sufficiently large scales to perform this test, while in the case of SKA-MID the stacked emission will be below the detection limit in both cases.

In addition to stacking clusters or filaments, there may be other ways to extract signals from the cosmic web \citep[][]{2011JApA...32..577B,va15survey}.  For example, recent statistical studies based on the cross-correlation between radio surveys and galaxy surveys  have derived  upper limits on the magnetization of filaments of the order of $\leq 0.1 \mu G$ , based on the absence of a strong correlation between synchrotron emission and the underlying galaxy distribution  \citep[][]{vern17,brown17}. Given the difficulty of  modeling the position-dependent contribution of galactic foregrounds and of resolved and unresolved radio galaxies to the weak large-scale structure emission, non-standard approaches to identify diffuse emission with a low signal-to-noise ratio might have to be adopted (e.g. based on Convolutional Neural Networks, Gheller et al., in prep.). 

\begin{figure}
\includegraphics[width=0.9\textwidth]{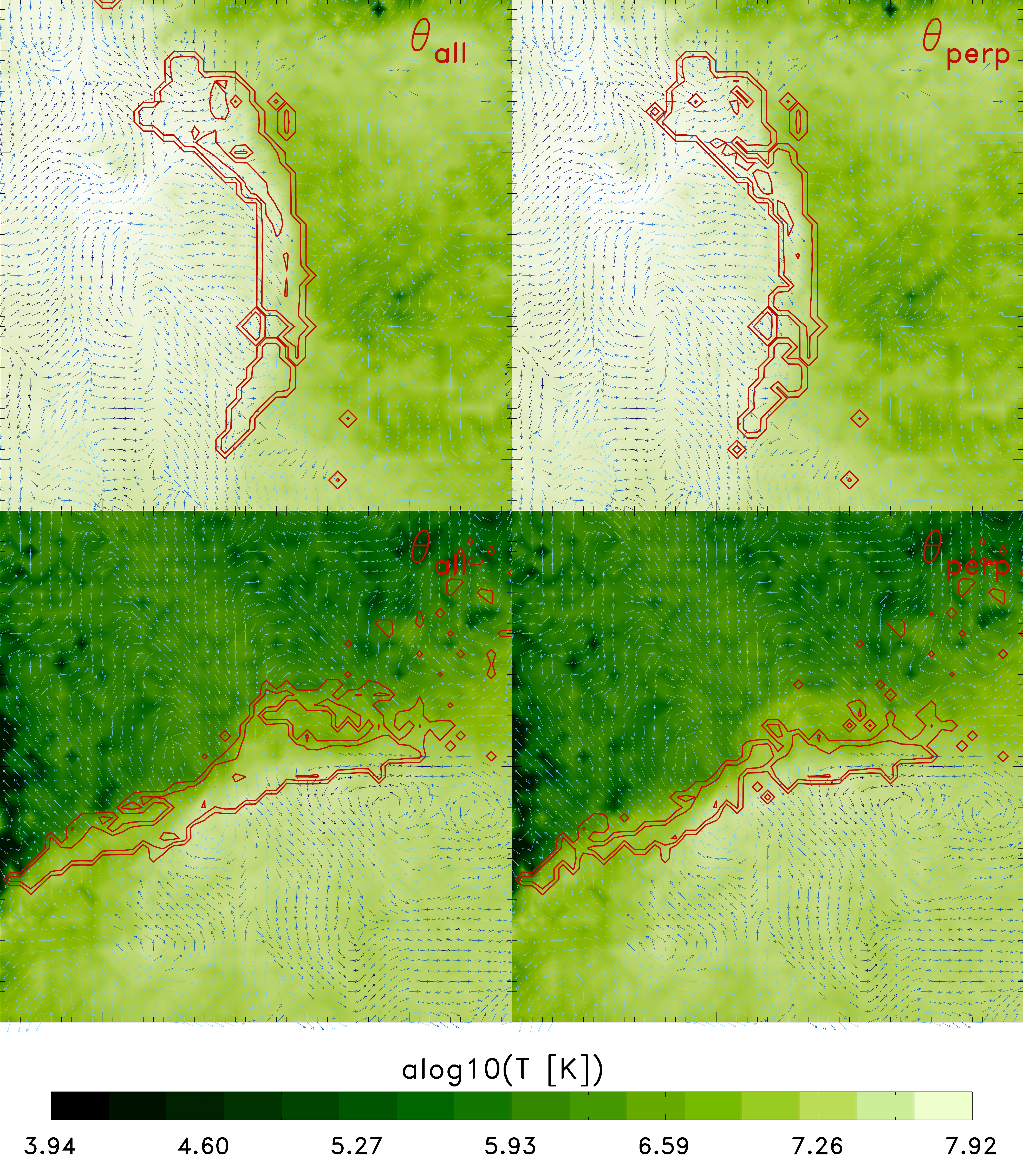}
\caption{The role of magnetic field-shock obliquity in the acceleration of cosmic ray electrons, for an internal merger shock leading to a radio relic (top panels) and for an outer accretion shock (lower panels). The red contours show the total radio emission, the green colors the gas temperature (see colorbar) and the arrows the orientation of the magnetic field (for clarity, the strength of each magnetic field vector has been normalized to one so that only the topology is concerned here. The left panels show the radio emission if shock acceleration of electrons only depends on the Mach number, while in the right panels the shock acceleration of electrons is limited to quasi-perpendicular, $\theta > 50^{\circ}$, shocks. This simulation employed adaptive mesh refinement down to a resolution of $36,6$ kpc/cell and a primordial model for magnetic fields (more details can be found in \citealt{wi17}). Each plot shows a region of the size $\sim 915 \ \mathrm{kpc}^2$.} 
\label{fig:wittor}
\end{figure}

\bigskip
Finally, the role of shock obliquity, defined as the angle $\theta$ between the shock normal and the up-stream magnetic field, in the acceleration of particles remains unclear. 
In the Earth's bow shock and in interplanetary shocks, electrons are {\it directly} observed to be accelerated to relativistic energies preferentially in quasi-perpendicular configurations, i.e. $\theta > 40-50^{\circ}$  \citep[e.g.][]{1999Ap&SS.264..481S,2006GeoRL..3324104O}. Recent particle-in-cell simulations of electron acceleration by weak intracluster shocks observed that only quasi-perpendicular shocks can pre-accelerate electrons, via shock-drift-acceleration (SDA), before being injected into DSA \citep[][]{guo14,guo14b}. 
With a higher-resolution version of clusters from the Chronos++ sample obtained with adaptive mesh refinement and starting from a simple primordial model, we have recently explored the link between the obliquity on scales resolved in the simulations and the acceleration of cosmic rays \citep[][]{wi16,wi17}. 
Fig.~\ref{fig:wittor} illustrates how switching off the acceleration of electrons for quasi-parallel shocks ($\theta \leq 50^{\circ}$) may change the observed synchrotron emission in internal (top) and external (bottom) accretion shocks  \citet{wi17}. The left panels show the radio emission if shock acceleration of electrons only depends on the Mach number (same synchrotron model by \citealt{hb07} as above), while in the right panels we limit the acceleration of electrons to  $\theta \geq 50^\circ$ shocks, based on \citet{guo14}. 
In typical merger shocks at low redshift the effect of an obliquity-dependent efficiency is small because the turbulent dynamics of the innermost ICM produces a distribution of angles  close to random ($N(\theta) \propto \sin{\theta}$).  Therefore, even by restricting the acceleration to quasi-perpendicular shocks, the radio emission is found to be reduced by just a factor  $\sim 2$. On the other hand, the radio emission from outer accretion shocks is more significantly reduced in the case of acceleration by quasi-perpendicular shocks only (lower panels):  not only the total power of the shock is significantly decreased  (i.e. by a factor $\sim 4-5$ in this case), but also the morphology of the emission becomes thinner.  

These results are consistent with the finding of \citet{wi17}, who found more parallel shocks compared to the random expectation, which occurs when large-scale primordial fields align with accretion shocks.

To complicate matters, \citet{2017ApJ...843..147M} recently detected significant electron acceleration by the $M_A \sim 10^2$ and $\beta \sim 10$ quasi-parallel shock, during the  crossing the Saturn bow shock by the Cassini space mission. These  plasma conditions are similar to the intracluster medium, and the acceleration may occur in the portion of the shock where upstream CR streaming instabilities  generate perpendicular small-scale magnetic field components at the shock surface. This could leading to particle acceleration  \citep[e.g.][]{2011ApJ...733...63R} and therefore stronger radio emission than in the case we just explored, with a suppression of acceleration for all shocks with $\theta < 50^{\circ}$ on $\sim 100 ~\rm kpc$ scales.

\begin{figure}
\includegraphics[width=0.99\textwidth]{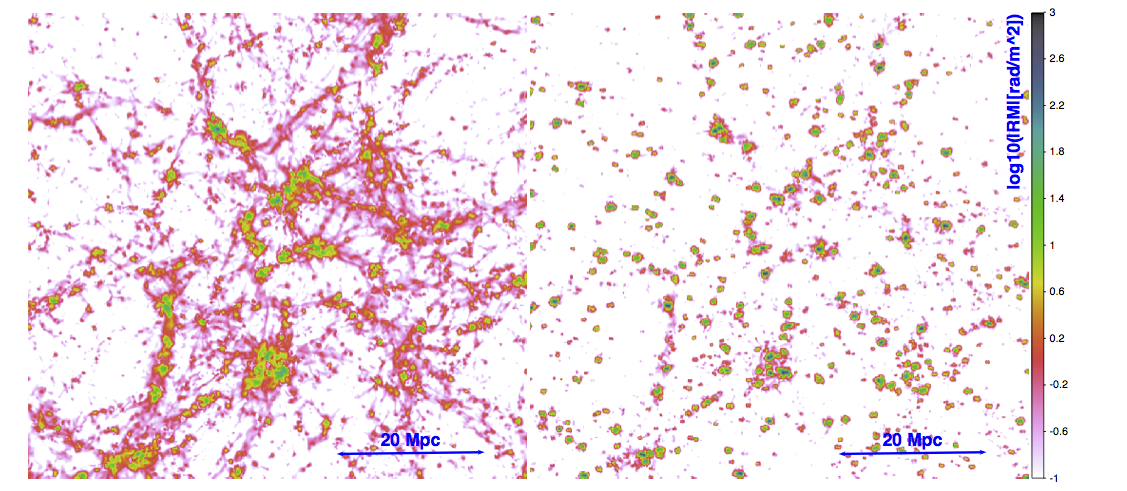}
\caption{Faraday Rotation measure from the same region as in Fig.~\ref{fig:map2} for our primordial (left) and AGN seeding scenario (right) from our  $2400^3$ runs at $z=0$. The line of sight considered here is $200 ~\rm Mpc$ and the closest structures are at a luminosity distance $d_L=100 ~\rm Mpc$ ($z \approx 0.024$).}
\label{fig:map_RM}
\end{figure}   

\begin{figure}
\includegraphics[width=0.95\textwidth]{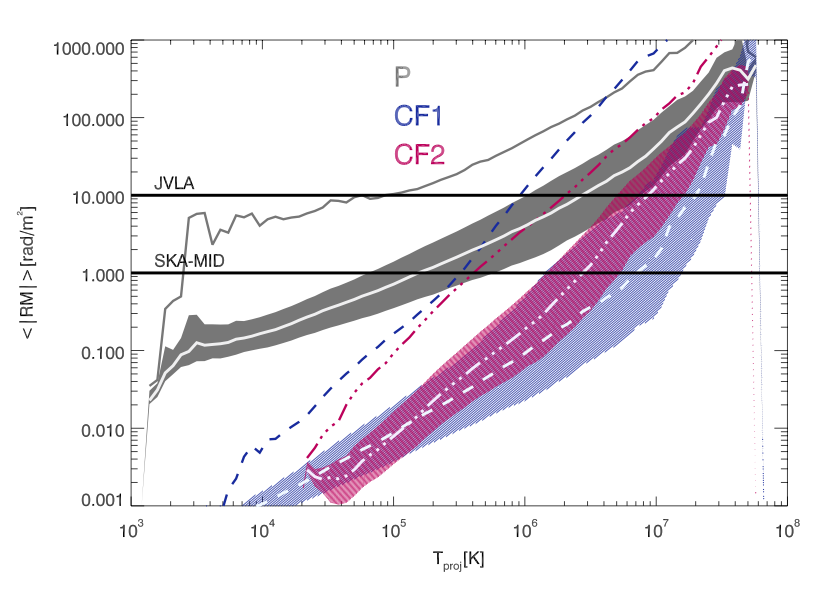}
\caption{Average RM as a function of average mass-weighted gas temperature in our large $2400^3$ runs at $z=0$, for a primordial or instead for the two astrophysical scenarios. The colored areas show the 25-75 percentile range, the white lines show the median for each model, while the upper lines show the 99 percentile of each model.  The horizontal lines show the approximate accuracy on RM which can be presently achieved with the JVLA, and the expectation for the SKA-MID. } 
\label{fig:RM_Temp}
\end{figure}   

\subsection{Faraday Rotation (in filaments and cluster outskirts)}
\label{rm}

The three-dimensional  distribution of magnetic fields in large-scale structures can be probed by 
Faraday Rotation measurement (RM)  of the emission from background polarized sources.
The signal goes as
\begin{equation}
  \rm RM[rad/m^2]=812 \int_0^{L} \frac{B_{\rm ||}}{\rm \mu G} \cdot \frac{n_e}{\rm cm^3} \frac{dl}{kpc},
\end{equation}
where $\rm ||$ denotes the component parallel to the direction of integration \citep[e.g.][]{ct02,mu04}. For a few decades, the study of RM from galaxy clusters has been essential to infer the tangled structure of intracluster fields \citep[][]{2003A&A...412..373V,mu04,gu08,bo10,vacca10}. 
So far, the RM of polarized sources shining through the cosmic web 
has been  carried out with statistical studies on the heterogeneous output of radio surveys 
 \citep[][]{Blasi.Burles..1999,2015aska.confE..92J,2016PhRvL.116s1302P,2016A&A...596A..22B}. 
For example, \cite{2016PhRvL.116s1302P}  observed that the RM  from distant radio sources do not exhibit any trend with redshift, while the extragalactic contribution  grows with distance, and from this they propose to constrain the amplitude of magnetic fields below $1.7 ~\rm nG$ within the gas Jeans length of structures.  However, the reliability of these estimates is limited by the fact that they only use two close frequencies in the NVSS survey \citep[][]{1998AJ....115.1693C}, thereby being prone to the $2 \pi$ ambiguity in the estimate of the rotation angle, and the effect of beam depolarization is not taken into account for the the $ 45''$ beam of NVSS.
So far, the study of RM in the outer regions of nearby galaxy clusters has  been limited to the prominent case of the Coma cluster \citet{bo13}. Using the RM of seven polarized sources in the direction of  the Coma radio relic and of an ongoing accretion from a possible intracluster filament, \citet{bo13}
 inferred a magnetic field of $\sim 1-2 ~\mu G$ close to the virial radius of the Coma cluster, significantly higher than what predicted by extrapolating the previous trend of RM from  sources observed in innermost $\rm Mpc^3$ volume of  Coma \citep[][]{bo10}. 
 
In the near future, the SKA-MID should be able to detect tiny differences in the magnetic field properties of the ICM, which are far beyond the capabilities of the present instruments, allowing the detection of hundreds of sources in the background of massive galaxy clusters \citep[][]{2013A&A...554A.102G,2015aska.confE..92J,2015aska.confE.105G,bo15}. Statistical methods based on Bayesian inference are also being developed to allow a robust removal of the various galactic and extragalactic contribution to the observed RM \citep[][]{2016A&A...591A..13V}. 
However, the RM data from the innermost cluster regions are not expected to teach us much about the origin of magnetic fields. This is because the small-scale dynamo amplification developed inside the inner cluster regions should be very efficient in removing any memory of seed fields as the bulk of observed magnetic energy is extracted from the cluster kinetic energy reservoir \citep[e.g.][]{cho14,bm16}. For example, \citet[][]{donn09} reported basically identical RM profiles of clusters simulated starting from competing seeding scenarios. 

With the  Chronos++ suite we can predict how RM should behave moving into the low density cosmic environment and away from the center of galaxy clusters. In a recent work (Vazza et al., submitted) we estimated that for this MHD scheme a resolution of at least $\sim  4 ~\rm kpc$ is  necessary to properly resolve the small-scale dynamo up to its final non-linear stage in a Coma-like cluster. The requirements on resolution should be less stringent for the RM in the external regions of clusters and in filaments, at least unless a dynamo on very small scales is possible (which is presently excluded by our numerical tests in \citealt{va14mhd}). 
As in the previous case, we show here the extreme comparison between the primordial and the astrophysical seeding scenario in our $200^3 \rm Mpc^3$ volume, for exactly the same field of view of the synchrotron emission (Fig.~\ref{fig:map_RM}). The RM signal is similar inside dense structures, and gets significantly different from cluster outskirts and filaments: $|RM| \sim 1-10 \rm ~rad/m^2$ in filaments in the primordial case, and $\leq 1 \rm ~rad/m^2$ in the astrophysical scenario. 
Unfortunately, these levels of RM are close to the sensitivity of existing and future polarimetric observations. 
As for the synchrotron emission, we show in Fig.~\ref{fig:RM_Temp} the ($|RM|,T_{\rm proj}$) relation from our simulated sky models. Similar to the previous case, the primordial scenario yields higher values of RM at all temperatures, but unlike the synchrotron case this is also true for the peaks in the RM distribution at all temperatures. 
Again, we expect to be able to detect the peak of the RM distribution 
 for $T \geq 10^{6} \rm K$ with current instrumentation (e.g. the JVLA in the Figure), and in  $T \geq 10^{5} \rm K$ with the SKA-MID.
 
For example, the planned deep polarization survey with Band 2 and Band 3 with SKA-MID 
 is expected to detect between $\sim 300$ and  $\sim 1000$ polarised
sources per square degree at 1.4 GHz, with 0.1-1 mJy and 
$1.6"$  resolution \citep[e.g.][]{2015aska.confE..95B,2015arXiv150102298T}.  For a Coma-like  cluster, this translates into a RM of $\sim 50$ for  polarized background sources, with a formal error in RM of the order of $\sim 1 ~\rm rad/m^2$.  Also before the advent of SKA, deep surveys in polarisation with ASKAP (Possum survey) and Meerkat (Mightee-Pol survey) are expected to enable a first testing of extreme scenarios of magnetogenesis.

\begin{figure}
\includegraphics[width=0.95\textwidth]{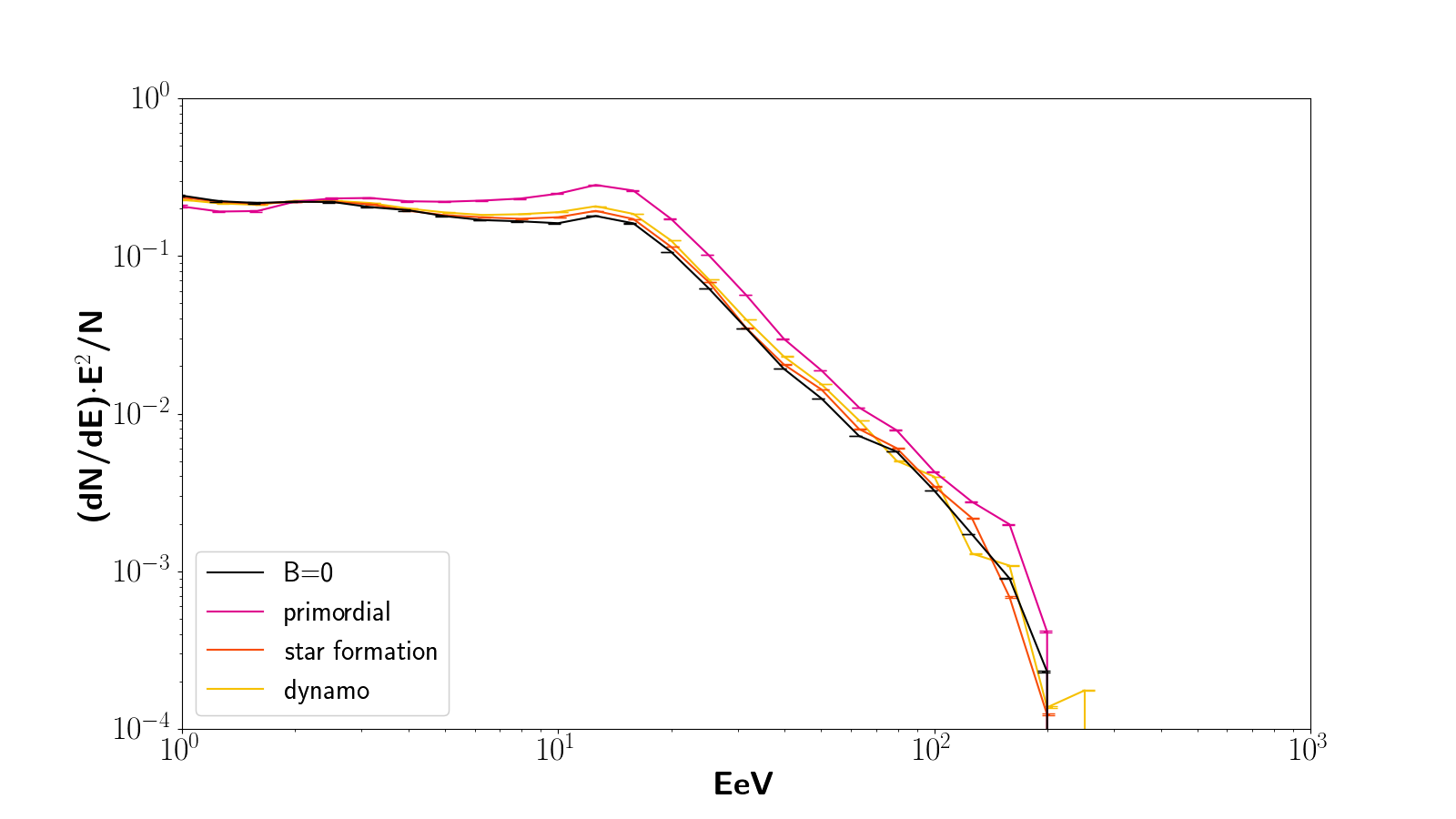}
\includegraphics[width=0.95\textwidth]{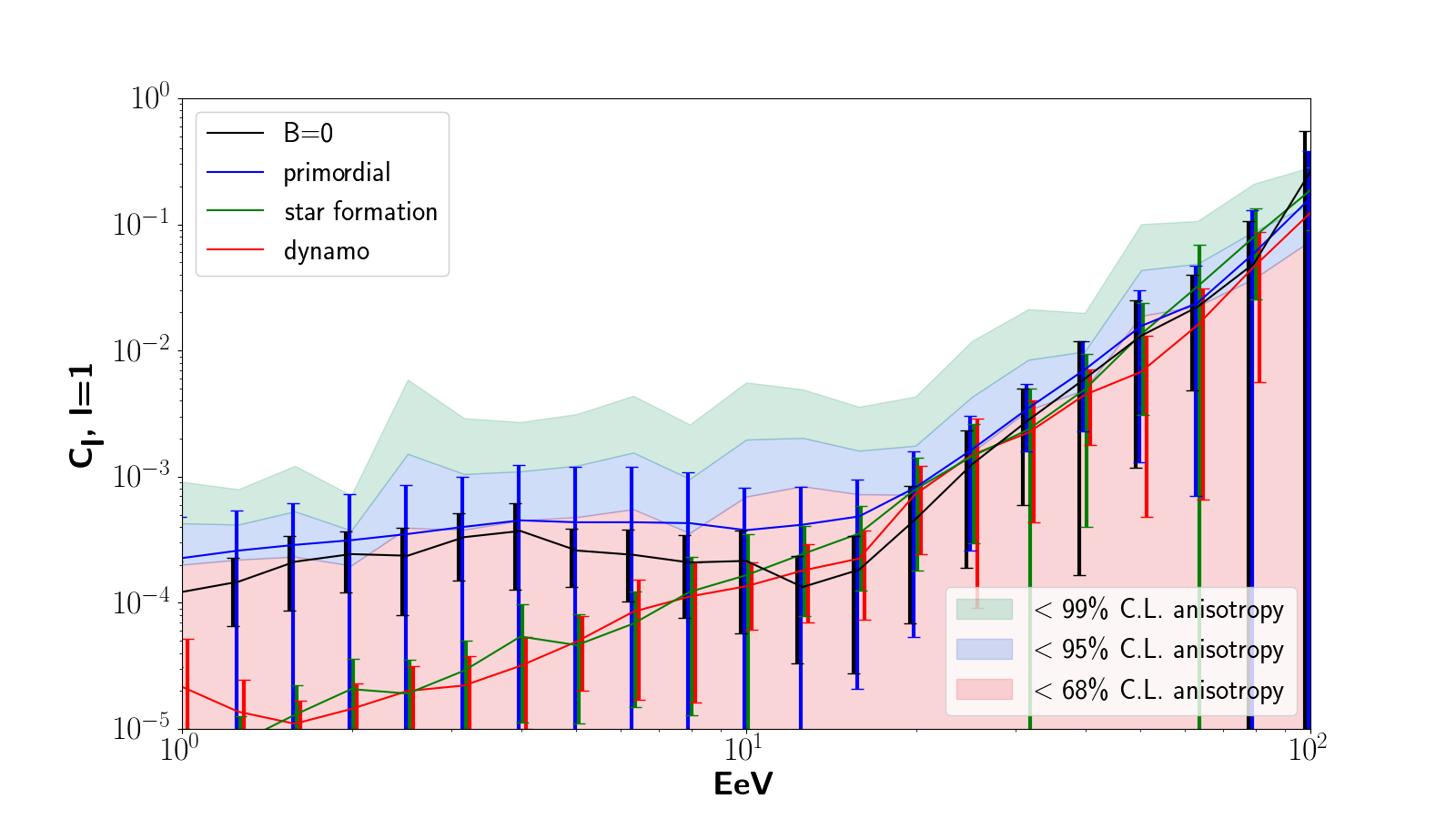}
\caption{Top: observed spectrum of UHECRs (averaged over 10 random observers) for different models: primordial, dynamo and astrophysical. Bottom: amplitude of the dipole in the spherical decomposition of full-sky maps of events for the same models. The colored regions show the 1-2-3 $\sigma$ deviations from the isotropic predictions for the same volume (based on the $B=0$ model with isotropic distribution of sources and same input spectra as the other models). The error bars show the $\pm 1 \sigma$ standard deviation in the population of 10 random observers. }
\label{fig:uhecrs}
\end{figure}

\subsection{High-energy physics probing large scale fields (in filaments and voids)}
 
Although voids are the most challenging cosmic environment for the study of cosmic magnetism due to the drop of all emission mechanisms there, they constitute the
environment where the largest differences between the various models is expected.  Our simulations show extremes ranging from $B_{\rm up} \sim 10^{-10} \rm ~G$ (primordial models) to $B_{\rm low} \sim 10^{-17} \rm ~G$. The first limit follows from the constraints from CMB analysis  \citep[][]{2010PhRvD..82h3005K,PLANCK2015}{\footnote{The analysis of higher-order statistics in the PLANCK data has been used to infer even lower limits, $\leq 0.01 ~\rm nG$, even if these estimates are more uncertain \citep[][]{2014PhRvD..89d3523T}.}}, while the lower limit is consistent with that $\geq 10^{-16} \rm ~G$ derived from the spectra of high-z blazar sources \citep[e.g.][]{2010Sci...328...73N,2014ApJ...796...18A,2015PhRvD..91l3514C}. {\footnote{See however \citet{2012ApJ...752...22B} for a different interpretation of these results.}}\\

Several interesting high-energy  phenomena may offer clues about magnetic fields on cosmological scales, dominated by voids and hardly affected by the details of galaxy formation. 

\subsubsection{Ultra-high Energy Cosmic Rays (UHECR)}
\label{uhecrs}
Cosmic rays with energies $\geq 10^{18} ~\rm EeV$ are thought to have a predominantly extra-galactic origin.  Their propagation towards Earth should be deflected by the Lorentz force of intervening magnetic fields. \citep[e.g.][]{Sigl:2003ay,2005JCAP...01..009D}. Large deflection angles might hamper the possibility of  locating the real sources of UHECRs at the highest energies, which is one of the main unknowns in this field of research  \citep[e.g.][]{2016arXiv161000944B}, but the deflections quoted in the literature greatly vary depending on the numerical methods and seeding models, ranging from $\leq 1^{\circ}$ to $\sim 30^{\circ}$ \citep[e.g.][]{2005JCAP...01..009D, 2014APh....53..120B,2014JCAP...11..031A}. 
With recent work  \citep[][]{hack16} we studied the propagation of UHECRs in {\enzo}-MHD simulations with the CRPropa3 code \citep[][]{2016JCAP...05..038A}, finding that the observed level of isotropy of $\geq 10^{18} \rm eV$ UHECR events  \citep[][]{2016APh....80..131A} can be used to limit the magnetic field in voids,  which must be $\leq 10 ~\rm nG$ \citep[][]{2017arXiv171001353H}. However, for lower magnetic fields the differences in the spectrum and in the spatial distribution of UHECRs between primordial and astrophysical seeding models are hard to tell. 

Here, we present a more recent analysis based on three resimulations of the (85 Mpc)$^3$ simulation, which have a better resolution and subgrid physics compared to our previous work. 
In detail, we used CRPropa3 to inject $10^8$ iron particles with random momentum in the $E=10^{18}-10^{21} ~\rm ev$  energy range, and we integrated their propagation until they were observed by 10 observers (located at the  centre of $\sim 10^{12}M_{\odot}$ halos, randomly chosen in the box). The evolution of the trajectory of each cosmic ray included the deflection  by the Lorentz force of  magnetic fields, energy losses, nuclear decay and photo-disintegration into secondary particles, and we assumed period boundary conditions in the computational volume. 
Since the distribution of the sources of UHECRs is unknown, as in \citet{hack16} we bracketed uncertainties with two injection scenarios:  a random distribution of sources or a scenario in which the probability of any cell to inject a cosmic ray is an increasing function of the gas density, i.e. UHECRs are preferentially originate from within large-scale structures. Here we present the results only for the latter scenario, noting we did not find large differences between the two. 
The above recipes were applied to three representative cases of the Chronos++ suite: the simplest primordial model, a dynamo model and a model with seeding from star winds (with the largest efficiency). 
 
The left panel of Fig.~\ref{fig:uhecrs} shows the arrival spectrum for each model, averaged over our 10 random observers. Within the error bars set by cosmic variance and number statistics of our simulations, we measure no significant difference in the received energy spectra of particles across most of the $E=10^{18}-10^{21} ~\rm eV$ energy range, also when comparing with the control $B=0$ case without magnetic fields (i.e. the UHECRs only propagate in a ballistic way). Minor differences are  found in the very high energy regime  (a few $10^{20} ~\rm eV$), where the primordial model have slightly flatter spectrum, which stems from the enhanced capture of high-energy cosmic rays around observers, whose halos are typically more magnetized than in the other scenarios. However, the difference will be hard to tell with any new observation, because the number statistics of these events is extremely small, i.e less than $\sim 1 \rm event/ (\rm km^2 100 yr)$.  
 
For each model and observer we also generated full-sky maps of events and computed the spherical harmonics (we refer the reader to \citealt{hack16} for the details on the procedure) to look for signatures of anisotropies at different energies.  The angular spectra have been averaged over observers, and compared to the $B=0$ case with an isotropic distribution of sources. 
The right panel of Fig.~\ref{fig:uhecrs} shows the amplitude of the dipole ($l=1$ harmonic) moment as a function of the arrival energy of UHECRs. 
Again, the largest departure from isotropy is observed for the primordial run, whose $l=1$ signal is everywhere higher (at $3 \sigma$) than the amplitude observed in the dynamo or astrophysical models. The effect is maximal at the lowest energies, $E \leq 10^{19} ~\rm eV$ events, highlighting the possibility of a higher deflection of UHECRs around structures closer to the observer. These results are explained by the typically larger magnetic fields found close to the observers ($\leq 500 ~\rm kpc$) in the astrophysical scenario (and partly in the dynamo one) that lead to a more isotropic distribution. Moreover, the larger fields typically found on $\geq \rm 1-10$ Mpc scales in the primordial scenarios may impose significant anisotropies on $ \geq 10^{\circ}$ scales. However, the amplitude of this effect is still compatible at $\sim 2-3 ~\sigma$ with random Poisson variance of an isotropic distribution of events of the baseline $B=0$ model, and is also within the present low bounds on the anisotropy observed with the Auger Telescope \citep{2016APh....80..131A} {\footnote{At a later submission stage of this work, \citet{2017arXiv170907321T} reported the detection of a significant anisotropy in the dipole distribution of UHECRs observed by Auger at $\sim 8 \cdot 10^{18} ~\rm eV$. Such a large anisotropy is hard to reconcile with any of our models, and it might be explained by a significantly  dipolar distribution of sources, as we discuss in detail in \citet{2017arXiv171001353H}.}. 
In summary, although we observe some minimal differences in the arrival spectra and angular distribution of UHECRs for different magnetization models, these differences are smaller than the effects caused by the distribution of sources of UHECRs, and by the scatter among different observers.  The latter can be mitigated by resorting to constrained simulations of the Local Universe \citep[][]{2017arXiv171001353H}.

\subsubsection{Fast radio bursts}
\label{frb}

Fast Radio Bursts (FRBs) can be used to probe extragalactic magnetic fields across cosmological distances.
So far, only a few dozen FRBs have been detected \citep[][]{frb_cat}, and the dispersion measure of their emission, defined as:
\begin{equation}
D = 1000 \int_0^z \frac{n_e(z)}{1+z} \frac{dl(z)}{dz} \rm pc/cm^3,
\end{equation}
(where $z$ is the redshift where the source is located) can give precious information on the intervening electron density along the line of sight,  $n_e(z)$.
In principle, the simultaneous detection of RM and $D$  for the same FRBs would allow deriving the properties of the intergalactic matter and of magnetic fields along the line of sight, over cosmological distances \citep[e.g.][]{2014ApJ...790..123A,2015MNRAS.451.4277D}.
However, so far it has been possible to detect RM only for a very few  FRBs, 
also with unexpectedly low measurements \citep[][]{2017MNRAS.469.4465P}. However, the situation is expected to improve thanks to ongoing  \citep[][]{2017arXiv170604459K} or future \citep[e.g.][ see also ALERT survey with Aperitif]{2013ApJ...776L..16T} radio surveys dedicated to detecting FRBs, with a few expected detections {\it per day}.  Compared to the standard analysis of RM from polarized radio galaxies, FRB may allow to probe lower values of RM, given their larger brightness $\sim \rm Jy$ flux at $1.4 ~\rm GHz$, which should allow to probe the rotation of their polarization angle with a typically higher accuracy. 

The largest boxes in the Chronos++ suite can test how different models for magnetogenesis can affect the signal from FRBs at  cosmological distances.
In Fig.~\ref{fig:frb}, we show the progression of dispersion measure and RM for a FRBs assumed to shine at $z=1$ and to be observed at $z=0$, in our three resimulations of the $200^3 ~\rm Mpc^3$ volume. Here we considered a random line of sight, obtained by pasting together 17  beams of one cell thickness crossing our volume at random positions and at different simulated redshifts; the combination of lines is exactly the same in the three resimulations, hence the observed differences in dispersion measure and RM are solely due to the different prescriptions for gas physics and magnetic fields. 
While the dispersion measure is only affected by gas physics in a minor way (due to the presence of denser clumps in the low-power AGN model), the observed trend of RM are hugely dependent on the assumed magnetic field model, with a  difference up to a factor $\sim 50$ in the final RM signal, when the primordial model (blue) or the high-power AGN model are compared. A large fraction of the RM signal from such a large integration volume comes from filaments and medium overdensity structures, in which cases the discrepancy among models is $\geq 10-100$ in RM (see Fig.~\ref{fig:RM_Temp}). In general, the integration up to large redshift may even increase the differences seen at low-z, due to the fact that the seeding from AGN could only magnetize a smaller fraction of the cosmic volume at earlier redshift. 
Different magnetogenesis models may produce differences in the RM but not in dispersion measure, and therefore the use of the dispersion measure-RM relation from observations \citep[e.g.][]{2016ApJ...824..105A} must be calibrated for different magnetic models before making claims about extragalactic magnetic fields. More work on the use of FRBs to study extragalactic magnetic fields is in preparation (Hinz et al., in preparation).

\begin{figure}
\includegraphics[width=0.95\textwidth]{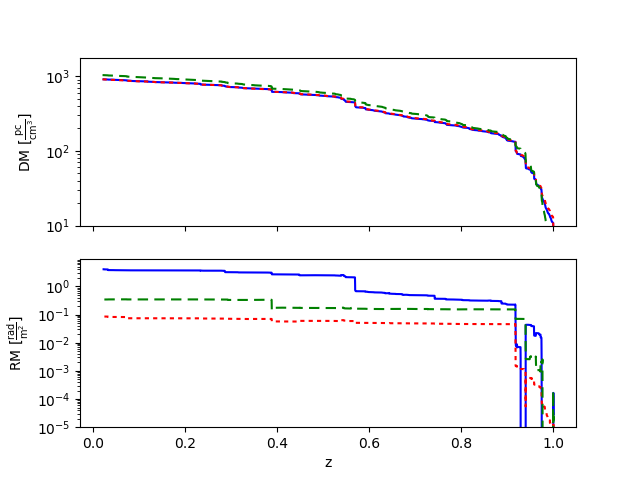}
\caption{Progression with redshift of the integrated dispersion measure (top) and RM (bottom) for a random line of sight from $z=1$ to $z=0$ in our largest simulations, comparing the primordial scenario (blue) with the low-power AGN (green) or the high-power AGN (red) scenarios for magnetogenesis.}
\label{fig:frb}
\end{figure}   

\subsubsection{Blazar emission and Axionlike particles}
\label{alps}

The magnetic fields in voids are key to understand the unexpected emission properties of $\sim ~\rm TeV$ blazars at high redshifts.  The high-energy, $\geq 100 ~\rm GeV$ emission from AGN is attenuated due to the interaction of  $\gamma$-rays with photons of the extragalactic background light. The produced pairs can inverse-Compton (IC)
scatter photons of the CMB and of the extragalactic background light and induce an
electromagnetic cascade \citep[][]{1994ApJ...423L...5A}, where the
pairs are deflected in external magnetic fields \citep[e.g.][]{2002ApJ...580L...7D,2010Sci...328...73N,2011A&A...529A.144T}. 
The lack of detected IC cascade component at $\sim$ GeV energies is probing the magnetic fields in the intervening cosmic volume (mostly dominated by voids)
and/or the filling factor of magnetic fields along the line of sight. From this effect, lower limits of $\geq 10^{-16} ~ \rm G$ on $\sim \rm ~Mpc$ scales have been derived \citep[][]{2009ApJ...703.1078D,2010Sci...328...73N,2011ApJ...727L...4D,2014ApJ...796...18A,2015PhRvD..91l3514C,2015PhRvL.115u1103C} 
The Cherenkov Telescope Array (CTA) will have the potential to further limit large-scale magnetic fields through this effect
\citep[][]{2013IAUS..294..459S,2016ApJ...827..147M}.

Finally, the presence of significant large-scale magnetic fields has  been suggested as a possible explanation for the puzzling lack of infrared absorption in the observed spectra of distant blazars \citep[e.g.][]{2012PhRvD..86h5036T,2012PhRvD..86g5024H}.  
Axion-like particles (ALPs)  are promising candidate for DM \citep[][]{1988PhRvD..37.1237R,2003JCAP...05..005C} and they can oscillate into high-energy photons (and back) in the presence of background magnetic fields. According to standard theory, $\gamma$-ray photons emitted by blazars must  be attenuated due to the production of pairs while interacting with the cosmic infrared background. However, 
this opacity may be reduced if $\gamma$-ray  photons propagate in external magnetic fields and therewhile oscillate into ALPs. Only a fraction of them is finally transformed back into photons near the observers. 
The first study of ALP oscillations used analytical prescriptions for the magnetic field distributions in galaxy clusters and voids \citep[][]{2012PhRvD..86g5024H}. In recent work based on the primordial $2400^3$ run of the Chronos++ suite, we simulated the propagation of photons from redshift $z=1$ and computed the expected conversion into  ALPs \citep[][]{2017PhRvL.119j1101M}.
This analysis showed that photons-ALPs oscillations are possible for lines of sight crossing structures with $\sim 1-10 ~\rm nG$ on scales of a few $\sim \rm Mpc$. Compared to a more idealized Gaussian distribution of extragalactic magnetic fields, the photons-ALPs conversions is found to cause a significant spectral hardening for emitted photons in the $\sim \rm ~ TeV$ energy range, and this effect can be measured with the upcoming CTA \citep[][]{2017PhRvL.119j1101M}.  If confirmed, this could be a powerful tool to perform a tomography of the magnetized cosmic web with ALPs, and to constrain the still larger range of parameters describing the mass and the coupling strength of these particles  \citep[][]{2014JCAP...12..016M}.
With future work, we plan to extend the analysis to competing mechanism for the seeding of large-scale fields, with the goal of assessing whether this effect vanishes for astrophysical scenarios of magnetic fields (Montanino et al., in preparation).

\begin{figure}
\includegraphics[width=0.95\textwidth]{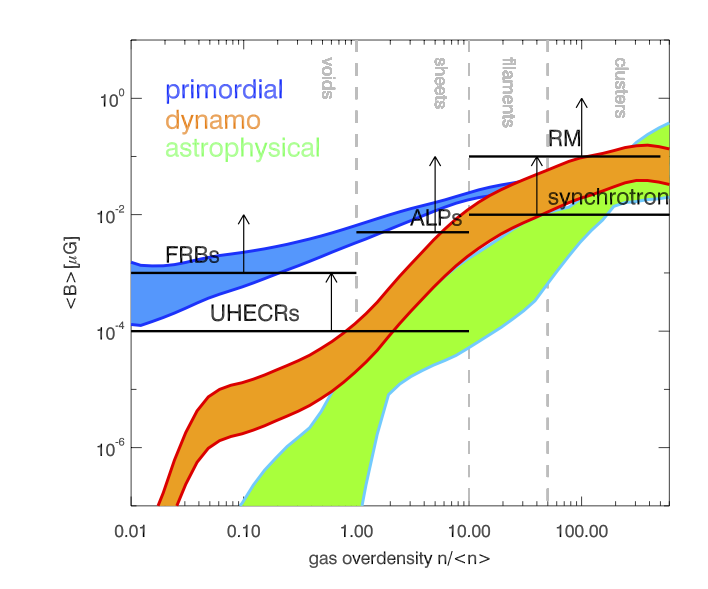}
\caption{Overview of the distribution of extragalactic magnetic fields predicted by our simulations and of the approximate regime where different observational effects can measure them. The colored area for each set of model  brackets the range of predictions for magnetic fields given already in Fig.~\ref{fig:phase1}.  }
\label{fig:final}

\end{figure}   
\section{Conclusions}
\label{conclusion}

In this work we introduced  a dedicated suite of cosmological magneto-hydrodynamical simulations with the {\enzo} code, that can be used to predict the evolution and present-day distribution of extragalactic magnetic fields, and also to assess the observable signatures associated to each of them.

In general, we find that if the present day magnetic fields observed in galaxies and clusters/groups of galaxies have a primordial origin from $\sim \rm nG$ magnetic fields, the magnetization in voids and in most filaments will always be larger than in any other scenario. Magnetic fields as large as $\sim 0.1 ~\rm \mu G$ may be found on large scales within filaments and in cluster outskirts only if the dynamo from injected solenoidal turbulence (which is still a small fraction of random kinetic energy in simulated large-scale structures) is efficient enough, with conversion efficiencies between random kinetic and magnetic energy of a few percent. Conversely, if the origin of magnetic fields is only ascribed to magnetized outflows powered by star formation winds and/or AGN, we expect a sharp drop in magnetization outside the virial radius of halos, and orders of magnitude lower magnetic fields in voids and in most of the volume of filaments.

Fig.~\ref{fig:final} gives a schematic overview of the range of uncertainties in the different seeding modes (the width of every group of models bracket the range of median in the ($B,n/\langle n \rangle$) relation of Sec.~\ref{3d}), as well as of the approximate overdensity and magnetic fields to which the observational probes is sensitive to.  

Hence, we predict for the observational signatures of different models: 

\begin{itemize}
\item {\it synchrotron emission from shock-accelerated electrons}: little to no emission should be detectable if fields were only injected by galaxy formation processes, except in the close proximity ($\leq \rm Mpc$ of active galaxies). The largest chances of detection (feasible in the ongoing low-frequency radio surveys with LOFAR, MWA, SKA-LOW) are for primordial scenarios starting from $\sim 0.1-1 ~\rm nG$. Present uncertainties in the details of shock-acceleration and in the role of shock obliquity in accelerating particles may make the above picture more uncertain.

\item {\it Faraday Rotation}: the differences between extreme models (e.g. high primordial scenarios vs astrophysical ones) can be detectable already at the periphery of clusters ($\geq R_{\rm 200}$) where we expect differences of order $\geq 10$ in the average RM of sources. A few tens of polarized background sources need to be observed for a few clusters before allowing a clear distinction of models; this should be feasible by the polarization survey planned for the SKA-MID. 

\item {\it Propagation of UHECRs:} the largest signatures are expected for primordial models starting from $\sim 0.1-1 ~\rm nG$ fields, and should result in a larger UHECR anisotropy and in a flatter spectrum of particles $\geq 10^{20}~ \rm eV$. However, in reality the uncertainties in the UHECRs' sources, as well as their composition will swamp all information about cosmic magnetogenesis. 

\item {\it Fast radio bursts:} we observe very significant (factor $\geq 10$) differences in the Rotation Measure of FRBs at $z=1$ when primordial and astrophysical models are compared. Most of the difference is produced by the integrated effect of voids, sheets and filaments. In combination with the dispersion measure of the burst, one can probe cosmological magnetic fields. While the number of FRBs with a known redshift as well as RM and dispersion measure is still too small, future surveys (e.g. "Alert" with Aperitif) will change this. 

\item {\it Other high-energy probes:} the conversion of $\gamma$-ray photons into ALPs as well as the production of an Inverse Compton Cascade around high-z blazars is expected to be maximal for primordial fields, owing to their enhanced efficiency in magnetizing voids.

\item The {\it synergy} between different observational techniques will be fundamental to remove the degeneracy between models: for example, while we expect a similar level of radio emission in primordial models starting from $\sim 1 ~\rm nG$ and in models where the magnetization of filaments comes from an efficient dynamo, the Faraday Rotation signal is expected to be larger and more structured in the second case. Likewise, the systematic  non-detection of Inverse Compton Cascade halos around blazars and of IR absorption hinting at ALPs oscillations may be used to constrain magnetic fields on $\sim \rm Gpc$ scales. 

\end{itemize}

In summary, we have presented an analysis of the impact of different models for  the origin of extragalactic magnetic fields. By coupling the evolution of galaxies and large-scale structures to the evolution of cosmic magnetism, we can enable different observational techniques to put competing scenarios of magnetogenesis to the test. In most scenarios, the periphery of clusters and cosmic filaments are the most interesting places where competing models can be tested. 

Given the planned capabilities of future radio surveys (most notably the ones that the Square Kilometer Array should deliver), the 
 most promising observational window into cosmic magnetogenesis is radio astronomy, in particular through the combination of observed synchrotron emission on large scales ($\geq ~\rm Mpc$) and deep polarization observations, able to measure the Faraday Rotation from sources probing the dilute gas in filaments and clusters outskirts. In voids, the high-energy sources at cosmological distances may constrain the residual magnetization from primordial seed fields. In all plausible scenarios, observations will need to be flanked by simulations in order to make statements about the origin of cosmic magnetism.

\section*{Acknowledgments}
The cosmological simulations described in this work were performed using the {\enzo} code (http://enzo-project.org), which is the product of a collaborative effort of scientists at many universities and national laboratories. We gratefully acknowledge the {\enzo} development group for providing extremely helpful and well-maintained on-line documentation and tutorials.\\
The computations presented in this work were mostly produced on Piz Daint (ETHZ-CSCS, Lugano) under allocations s585 and s701, and in small part at Juelich Supercomputing Centre (JSC), under projects no. 11823, 10755 and 9016.  F.V. acknowledges financial support from the grant VA 876-3/1 by DFG, from the European Union's Horizon 2020 research and innovation programme under the Marie-Sklodowska-Curie grant agreement no.664931,  and from the ERC Starting Grant "MAGCOW", no.714196.  SH acknowledges support by the Deutsche Forschungsgemeinschaft (DFG) through grant BR 2026/25.  DW acknowledges support by the DFG through grants SFB 676 and BR 2026/17.  We gratefully acknowledge A. Bonafede, G. Brunetti, D. Montanino, A. Mirizzi , M. Viel, C. Ferrari at T. Jones for ongoing collaborations and useful feedback on the several lines of research relevant for this work. 

\bibliographystyle{jphysicsB}
\bibliography{franco}

\appendix

\section{Simulated star formation and feedback}
\label{sfr}
The process of star formation in {\enzo} is implemented through star particles that are formed on the fly, according to several possible recipes \citep[][]{enzo14}. Stars return thermal energy, gas and metals back into the gas phase. At the resolution probed in our simulations (which is too coarse to properly resolve the multi-phase process of star formation) we form star particles based on a density criterion at a rate which depends on a fixed timescale \citep[][]{2003ApJ...590L...1K}. Central to this star formation model are the choice of a minimum gas density threshold, $n_{*}$ and the dynamical timescale $t_{*}$ and the minimum star mass, $m_{*}$. Whenever $n \geq n_{*}$, the gas is contracting ($\nabla \cdot \vec{v} < 0$), the local cooling time is smaller than the dynamical timescale ($t_{\rm cool} \leq t_*$) and the baryonic mass is larger than the minimum star mass ($m_b=\rho \Delta x^3 \geq m_*$ ) we form stars with a mass $m_* =m_b \Delta t/t_{\rm *}$  By testing with smaller runs, by removing the corresponding mass from the gas mass in the cell. This recipe has been proposed to reproduce the observed Kennicutt's law \citep[][]{1998ApJ...498..541K} and by benchmarking against the observed cosmic star formation history  we explored a few possible variations of the above parameters (see Tab.~2). The feedback from the star forming process (e.g. supernovae explosion) depends on the assumed fractions of energy/momentum/mass ejected per each formed star particles, $E_{SN}= \epsilon_{SF} m_* c^2$. We assume the feedback energy  to be released entirely in the thermal form (i.e. hot supernovae-driven winds), which typically turns into the formation of  pressure-driven winds around our halos ($v_{\rm wind} \sim 10-10^2$ km/s).  For the injection of magnetic fields by stars,  we additionally introduced the generation of magnetic dipoles during each feedback episode. The dipoles are randomly oriented along one of the coordinate axis of the simulation, and their total energy scales with the thermal feedback energy, $E_{\rm b,SN}=\epsilon_{\rm SF,b} \cdot E_{\rm SN}$, typically with $\epsilon_{\rm SF,b}~\sim 1-10 \%$.  A list of all parameters for star formation physics assumed in our runs is given in Tab.~2. 

For an overview of the performances and limitations of the recipes for star formation and feedback in {\enzo} we refer the reader to the recent survey of models by \citet{2013ApJ...763...38S} and \citet{2015ApJ...811...73L}.  We also notice that recent works explored the complexity of properly implementing supernova-driven feedback in simulations, in a way that can be independent on the assumed spatial resolution \citep[][]{hop17}; however these effects can only be studied in detail on a much smaller scale that what is accessible by our runs here. 

Fig.~\ref{fig:sfr} gives the cosmic star formation (SFR) history for a few representative runs from the Chronos++ suite.  The result from our runs is compared with the observational compilation of infrared and ultraviolet observations (taken from \citealt{2014ARA&A..52..415M}). The match is reasonably good for our CSF2, CSF3 (and additionally CSFBH5 and CSFBH4 which also contains supermassive black holes, see following Section), while  models CSF0 and CSF1 (and CSFBH1) present a marked drop in the SFR at late times. The combined effect of the limited spatial resolution and DM mass resolution in our runs ($\Delta x=83.3$ kpc and $m_{\rm DM}=6.19 \cdot 10^{7}M_{\odot}$, respectively)  is expected to quench the formation of galaxies $ \leq 10^8 M_{\odot}$, and therefore our simulations still lacks an important contributor to the cosmic SFR \citep[e.g.][]{2014MNRAS.445..175G}. However,  the contribution to both the chemical, thermal and magnetic enrichment of the intergalactic medium by galactic outflow should be largely dominated by $\sim 10^{10}-10^{11} M_{\odot}$ \citep[e.g.][]{sam17}, which are resolved in our runs. We also notice that the 
last stage of the wind developments in our runs should be powered by the non-thermal pressure of magnetic fields, as in cosmic-ray driven winds\citep[e.g.][]{2012A&A...540A..77D,2014MNRAS.437.3312S,sam17}, given the identical equation of state $P_{\rm cr} \propto n^{4/3}$ in the ultra-relativistic regime vs $P_{\rm B} \propto n^{4/3}$ in the frozen-field approximation. 

Fig.~\ref{fig:sfrB} shows the average magnetization output of several star formation and feedback recipes in the Chronos++ suite: the left panel show the average magnetic field injected as a function of time by star forming particles, averaged inside the cell volume ($\Delta x^3=83.3^3 ~\rm kpc^3$), while the right panel show the average magnetic field in the total volume (i.e. also by taking into account the number of star forming particles in each model). 
The largest magnetic fields at the injection ($\sim 3 ~\rm \mu G$) are found in the CSF3 and in the CSFBH5 abd CSFBH6 runs, owing to the largest efficiency assumed there,  while  $\sim 10^2$ lower values are found in less energetic scenarios. On average, on a random $\rm Mpc^3$ volume in the simulation we have $\sim 1-10 ~\rm nG$ of magnetic fields produced from star formation in the CSF3 model, which is in good agreement with the recent semi-analytical estimate by \citet{sam17}, while we measure orders of magnitude lower fields for our lowest efficiency models. While BH particles contribute to additional magnetization inside halos, the volume averaged magnetic field output of star forming particles in runs with BH is nearly the same as without them, because most of star formation happens in regions away from BH. 
We remark that the above values do not strictly represent a background value for the magnetic fields in voids, because the fields injected by star forming particles do not uniformly expand onto large scales on a finite timescale,  and are also further affected by $\propto n^{2/3}$ adiabatic losses while expanding from dense star forming regions. 

The effects of star formation (also combined with feedback from BH, see below) on the simulated clusters/groups in a few of our runs is shown in Fig.~\ref{fig:clusters}: the cooling of gas into stars and their feedback steepens the mass-temperature relation within $R_{\rm 500}$ towards the values observed for X-ray groups \citep[][]{2011A&A...535A...4R,2011A&A...535A.105E}, by increasing the typical temperature of $M_{\rm 500} \leq 10^{14}M_{\odot}$ systems. 

\begin{figure}
\includegraphics[width=0.9\textwidth]{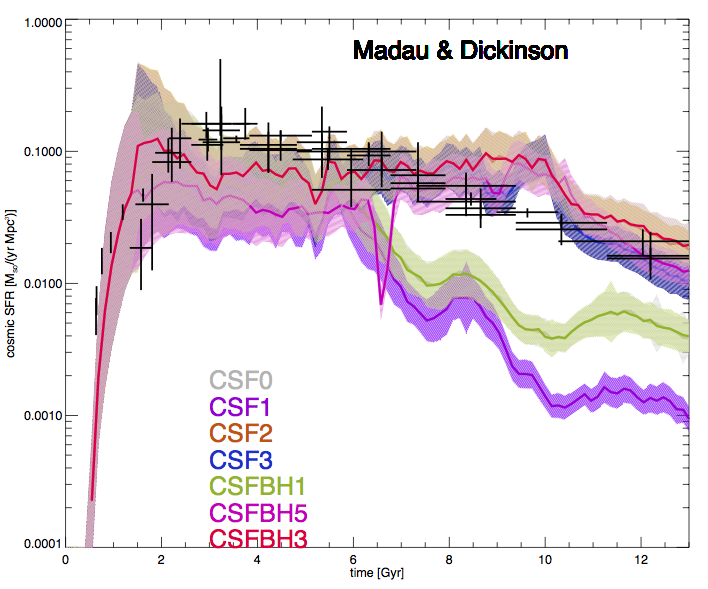}
\caption{Simulated cosmic star formation history for a few our runs (coloured lines with $\pm 3 \sigma$ variance) vs observed cosmic star formation history from the collection of observations in  \citet{2014ARA&A..52..415M} (grey points with errorbars).}
\label{fig:sfr}
\end{figure}

\begin{figure}
\includegraphics[width=0.95\textwidth]{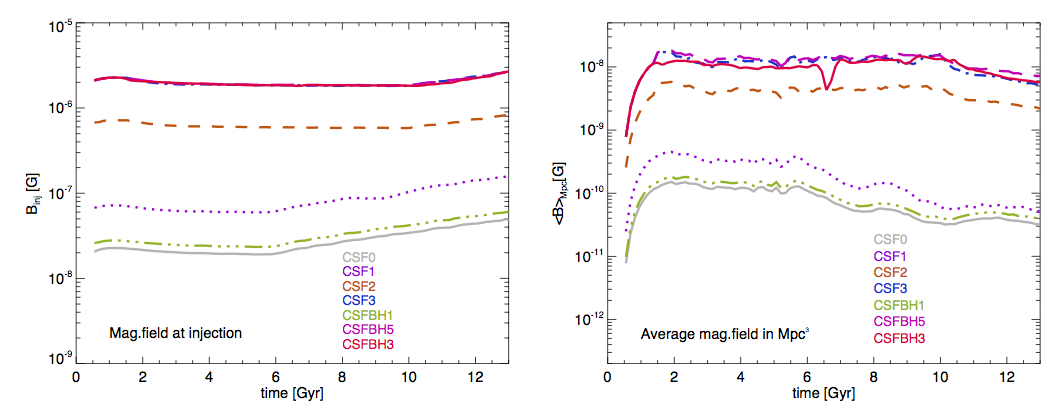}
\caption{Evolution of magnetic fields injected by star forming particles in the same runs of Fig.~\ref{fig:sfr}: the left panel shows the average magnetic field at the injection point of star forming particles, the right panel shows the average magnetic fields released per cubic Mpc$^3$ by all active star particles in the entire simulated (85 Mpc)$^3$.}
\label{fig:sfrB}
\end{figure}

\begin{figure}
\includegraphics[width=0.95\textwidth]{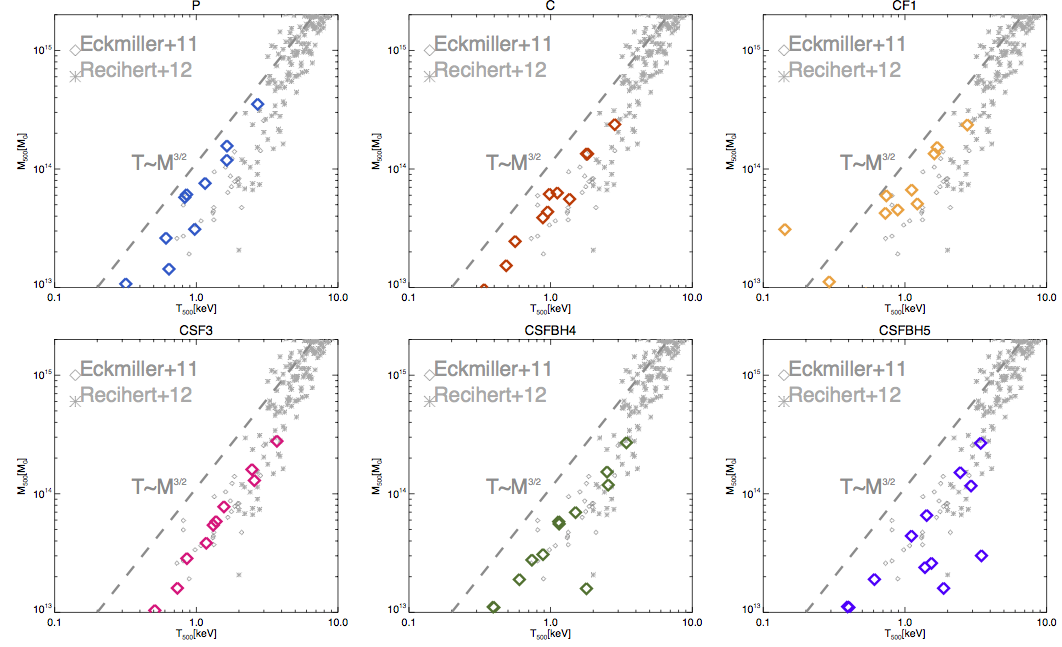}
\caption{Temperature-Mass relation  at $z=0$ within $R_{\rm 500}$ for clusters in a subset of the Chronos++ suite. The gray line shows the self-similar scaling relation and the additional gray points show the X-ray observations by \citet{2011A&A...535A.105E} and \citet{2011A&A...535A...4R}, for comparison.}
\label{fig:clusters}
\end{figure}

\section{Simulated supermassive black hole formation and feedback}
\label{bh}
In a few of our runs we included the growth of supermassive black holes and their thermal/magnetic feedback on the surrounding cosmic gas. Using existing implementations in {\enzo}, we inject black hole seed particles at  $z=4$, by locating them at the centre of all $\geq 10^{9} M_{\odot}$ halos identified in the simulation \citep[][]{2011ApJ...738...54K}. For each seed particle we assume a fixed initial mass of $M_{\rm BH}=10^4 M_{\odot}$.   Each BH particles accretes gas mass with assuming a Bondi-Hoyle rate, with the following ad-hoc numerical prescriptions to overcome the problems connected to the coarse resolution of our runs: a) we assumed the gas is accreted at the fixed temperature of $3 \cdot 10^5 \rm K$, because a precise estimate of gas temperature on the accretion radius is not feasible at our resolution; b) we boost the Bondi accretion rate by a factor $\alpha_{Bondi}=10^2-10^3$, to overcome the effect of the limited gas density which can be resolved around our BH particles. In a couple of runs, we instead assumed a fixed mass accretion rate onto BH particles, $10$ or $100$ $M_{\odot}/yr$. 
Each BH particle is allowed to release  thermal feedback on the surrounding gas, as an extra thermal energy output from each black hole particle, assuming an efficiency $\epsilon_{\rm BH} = \Delta M c^2 \Delta t/E_{\rm BH}$  between the accreted mass, $\Delta M$,  and the feedback energy, $E_{\rm BH}$ . As in the case of stellar feedback, with new code implementation we linked the release of feedback energy to the injection of magnetic fields from the BH region (in the form of magnetic dipoles), assuming $E_{b,AGN}= \epsilon_{b,AGN} E_{\rm BH}$, where we explored values in the range $\epsilon_{\rm b,AGN}=1-10 \%$. ) of the energy conversion into magnetic field dipoles at the jet base.  A list of all parameters for BH physics assumed in our runs is given in Tab.~2. 

Fig.\ref{fig:clusters} (lower panels) show the impact of feedback from BH particles in a few runs of the Chronos++ suite. The slightly larger scatter of the mass-temperature relation, compared to the case where only star formation and feedback is adopted, is a result of the intermittent feedback activity from AGNs, which can significantly heat the core regions of groups. 

With an alternative procedure designed  to spare computational resources in our largest $2400^3$ runs, we also used a much simpler recipe for AGN and stellar feedback, where for $z \leq 4$ we allowed for the impulsive injection of thermal and magnetic energy (also assuming a fixed efficiency, $\epsilon_{b}$), whenever the 
 the physical gas number density exceeded a critical value of $10^{-2} ~\rm cm^{-3}$ , with a fixed energy per event,  $E_{\rm BH}=10^{58}-5 \cdot 10^{59}$ erg . Our previous work \citep[][]{va16cr,hack16} has shown that this approximated approach is able to reproduce the observed steepening of the Temperature-Mass relation in galaxy groups (relative to the self-similar scaling) and to reasonably reproduce the observed profiles of gas density, temperature and pressure in galaxy clusters. 
 
For a complete overview of the implementation of BH particles and thermal/kinetic feedback in {\enzo} we refer the reader to the original work by \citet{2011ApJ...738...54K}. We also notice that recent works explored the complex interplay between nuclear star formation in galaxies and the growth of BH \citep[e.g.][]{2017MNRAS.469..295B}, as well as the effect of the detailed implementation of AGN-driven jets in the intracluster medium \citep[e.g.][]{2017arXiv170507900B}, but these effects can only be studied on a much smaller scale that what is accessible by our runs. 
 
\section{Zeldovich approximation for primordial magnetic seed fields}
\label{zeld}
A convenient way we explored to generate an initial distribution of magnetic fields correlated with initial density perturbations of large-scale structures is based on the Zeldovich approximation.   In order to ensure that $\nabla \cdot \vec{B} \equiv 0$ at the start, we   initialise the magnetic field vectors perpendicular to the 3-D gas velocity field computed with the Zeldovich approximation: in the Zeldovich approximation the velocity field is entirely compressive by definition,  hence a  field perpendicular to  it is a purely solenoidal. In order to match with the simpler  intialisation of  $\vec{B_0}$ in the other uniform approach,  the r.m.s values of each component generated with the Zeldovich approximation are renormalized within the cosmic volume, so that $\sqrt{|<B^2>|} = B_0$.  This initialisation strategy is convenient as it ensures that $\nabla \cdot \vec{B} \equiv 0$ by construction and without having to resort to cleaning schemes or higher order derivcatives in Fourier space, and moreover it by-passes the problem of generating very large initial distribution of magnetic fields via FFTs, which can be prohibitively large for the $1024^3-2400^3$ simulations we are concerned with.  In this project, we tested two different random orientations of the initial magnetic field vector produced with the Zeldovich approximation. 

\section{Subgrid dynamo model}
\label{sg}
Our  run-time implementation of a sub-grid (SG) model for dynamo growth of magnetic fields tries to estimate the level of dynamo amplification which may occur given the level of kinetic energy resolved in the simulation and associated to the solenoidal velocity component ($\nabla \cdot \vec{v}=0$), in the course of the root-grid timestep, $\Delta t$ (typically in the range of $10 ~\rm Myr$). The non-linear stage of dynamo growth in our runs may be hampered by the moderate  Reynolds number we can achieve ($R_e \leq 300$ in most of the volume, e.g. \citealt{va14mhd}) and therefore the dynamo amplification resolved in our runs is only a lower limit on the true amplification process. 
 The SG model proposed here  relies on measuring the gas vorticity at-run time, and in estimating the energy transfer between the cascading solenoidal kinetic energy (in a Kolmogorov model of turbulence) and the magnetic energy in the dynamo regime. 
 The solenoidal component  is well known to the be one connected to the triggering of magnetic dynamo \citep[e.g.][]{ry08} and the  vorticity squared (also known as ???enstrophy???, $(\nabla \times \vec{v})^2 \equiv \epsilon_{\omega}$) gives a convenient way of estimating the turbulent dissipation rate on the fly in our simulations as $F_{\rm turb} \simeq \eta_t \rho \epsilon_{\omega}^3/L$ , where $\eta_t=0.014$ and $L$ is the stencil to compute the vorticity \citep[][]{jones11,po15,va17turb,wi17b}.  The fraction of this turbulent kinetic power which is  ultimately converted into magnetic energy, $\epsilon_{\rm dyn}$, sets the amplified magnetic energy as $E_{\rm B,dyn} = \epsilon_{\rm dyn}(\mathcal{M})F_{\rm turb} \Delta t$. 
 While  direct numerical simulations of idealized intracluster medium dynamo suggest $\epsilon_{\rm dyn} \approx 0.05$   \citep[][]{bm16}, for our cosmological simulations we need a more general approach to link the amplified magnetic energy local plasma conditions across cosmic environments.
 Here we rely on the survey of simulations performed by  \citet{fed14}, who simulated small-scale dynamo in a variety of  conditions with different Mach numbers, $\mathcal{M}$,  for the forcing of turbulence.   Based on this prescription, we can estimate the saturation level and the typical growth time of magnetic fields as a function of the local Mach number of the flow, using the fitting formulas given by \citet{fed14}. In particular, $\epsilon_{\rm dyn}=\epsilon_{\rm d}(\mathcal{M})$ i.e. its value can be guessed from the fitting formulas once that the timestep and the local Mach numbers are known, as  $\epsilon_d(\mathcal{M}) \approx (E_B/E_k) \Gamma \Delta t$, where $E_B/E_k$ is the estimate ratio between magnetic and kinetic energy at saturation and $\Gamma$ is the typical growth rate, which we both take from \citet{fed14} as a function of the flow Mach number. Once we compute the amplified magnetic energy, $E_{\rm B,dyn}$, we use it to generate a new magnetic field vector, $\vec{\delta B}$,  which we  add to the already existing field, imposing that it is parallel to the local direction of the gas vorticity, so that the new generated field is also solenodial by construction. A corresponding amount of kinetic energy is removed from the cells, and momentum is removed assuming an isotropic dissipation of the small-scale velocity vectors.  This procedure is manifestly  simpler than more sophisticated SG models on the market, where the SG turbulent energy is measured via Favre filtering, and the electromotive force is added in a self-consistent way \citep{gr16}. However, our tests show that its application to simulated large-scale structures well matches the prediction that can be worked out in post-processing, based on the measured value of vorticity \citep[e.g.][]{ry08}.

\begin{table}
\begin{center}
\caption{Main parameters of runs in the Chronos++ suite of simulations using star forming particles and/or supermassive BH. See Table 1 in the main text for additional information.}
\small
\centering \tabcolsep 2pt
\begin{tabular}{c|c|c|c|c|c|c|c|c|c}
   $n_*$ & $t_*$ & $m_*$ & $\epsilon_{\rm SF}$ &$\epsilon_{\rm b,SF}$ & $\alpha_{\rm Bondi}$& $t_{\rm BH}$ & $\epsilon_{\rm BH}$& $\epsilon_{\rm b,BH}$ & ID \\ 
   
  $[1/cm^3]$ & [$yr]$ & $M_{\odot}$ &  &  &  &  $[yr]$  &   &  &  \\   \hline 
   $10^{-3}$ & 1.5 & $10^7$ & $10^{-8}$  & 0.01 & - &  - & - & - & CSF1\\
   $5 \cdot 10^{-4}$ & 1.5 & $10^7$ & $10^{-7}$ & 0.1  & - &  - & - & - & CSF2\\
   $5 \cdot 10^{-4}$ & 1.0 & $10^7$ & $10^{-6}$ & 0.1  & - &  - & - & - & CSF3\\
  $2 \cdot 10^{-4}$ & 1.0 & $10^7$ & $10^{-7}$  & 0.1  & - &  - & - & - & CSF4\\
  $2 \cdot 10^{-4}$ & 1.0 & $10^7$ & $10^{-6}$ & 0.1  & - &  - & - & - & CSF5\\
  $10^{-3}$ & 1.5 & $10^7$ & $10^{-8}$  &  0.01 & $10^2$ & $10^7$  & 0.05 & 0.01 & CSFBH1\\
  $10^{-3}$ & 1.5 & $10^7$ & $10^{-8}$  &  0.01 & $10^3$ & $10^7$  & 0.05 & 0.01 & CSFBH2\\
  $2 \cdot 10^{-4}$ & 1.0 & $10^7$ & $10^{-7}$  &  0.1 & $10^2$ & $10^7$  & 0.05 & 0.1 & CSFBH3\\
  $2 \cdot 10^{-4}$ & 1.0 & $10^7$ & $10^{-6}$  &  0.1 & $10^3$($10~M_{\odot}/yr$ fixed )& $10^7$  & 0.05 & 0.1 & CSFBH4\\
 $2 \cdot 10^{-4}$ & 1.0 & $10^7$ & $10^{-6}$  & 0.1 & $10^3$($100~M_{\odot}/yr$ fixed) & $10^7$  & 0.05 & 0.1 & CSFBH5\\
$2 \cdot 10^{-4}$ & 1.0 & $10^7$ & $10^{-6}$  & 0.1 & - & -  & - & - & CSFDYN1\\ 
   
  \end{tabular}
  \end{center}
\label{table:tab2}
\end{table}

\end{document}